\documentclass[trackchanges]{aastex7}
\usepackage{amsmath}
\usepackage{graphicx}
\usepackage{subcaption}

\begin{document}

\title{Dissecting the multiple-component outflow in NGC 5548 with absorption-line Variability}

\author{Yaqi Chen}
\email{}
\affiliation{Department of Astronomy, University of Science and Technology of China, Hefei, Anhui 230026, China
}%
\affiliation{School of Astronomy and Space Science, University of Science and Technology of China, Hefei, Anhui 230026, China
}

\correspondingauthor{Zhicheng He}
\email{zcho@ustc.edu.cn}

\author{Zhicheng He}
\email{}
\affiliation{Department of Astronomy, University of Science and Technology of China, 
 Hefei, Anhui 230026, China
}%
\affiliation{School of Astronomy and Space Science, University of Science and Technology of China, Hefei, Anhui 230026, China
}%

 \correspondingauthor{Guilin Liu}
\email{glliu@ustc.edu.cn}
\author{Guilin Liu}%
\email{}
\affiliation{Department of Astronomy, University of Science and Technology of China, 
 Hefei, Anhui 230026, China
}%
\affiliation{School of Astronomy and Space Science, University of Science and Technology of China, Hefei, Anhui 230026, China
}%

\begin{abstract}
AGN-driven outflows are routinely invoked as a key agent of supermassive black holes to regulate the evolution of galaxies. The radial distance from the central engine is a crucial parameter for evaluating the impact of these outflows on the host galaxy. In this work, we estimate the radial distances of ultraviolet (UV) outflow components in NGC 5548 using the most up-to-date absorption-line variability method, combined with multi-epoch HST/COS spectroscopy from the 2014 AGN STORM campaign and archival data observed in 2013. The recombination timescale ($t_r$) of the absorbers are measured by analyzing the detection rate curves of absorption-line variability. In particular, the detection rate curves of the absorption troughs showing blended multiple velocity components are featured by distinct ``multi-step'' profiles, allowing for measuring $t_r$ for individual components. Among the 6 identified outflow components, four are found to be a few pc from the center and two are 30-40 pc away. Our results agree well with the more reliable results in the literature on components 1 and 4, and show overall consistency with previous works, demonstrating the power of our new methodology especially when it is aided by densely sampled HST spectra.  

\end{abstract}


\section{Introduction}
Active galactic nuclei (AGNs) are among the most luminous and energetic phenomena in the universe, residing in the centers of galaxies and powered by accretion onto supermassive black holes (SMBHs) (\cite{1969Natur.223..690L,1984ARA&A..22..471R}).
As matter is funneled into the SMBH via an accretion disk, energy is released in the form of electromagnetic radiation, relativistic jets, and wide-angle, non-relativistic outflows.
This energy output couples with the interstellar medium (ISM) of the host galaxy, clearing or heating the surrounding gas to suppress or enhance star formation, and driving the observed correlations between central supermassive black holes and the stellar components of their host galaxies. Such processes, collectively termed active galactic nucleus (AGN) feedback, are considered fundamental to galaxy formation and evolution. (e.g., \cite{2005MNRAS.361..776S,2006ApJS..163....1H,2006MNRAS.370..645B,2007ApJ...665..120A,2008MNRAS.390.1399B,2008MNRAS.391..481S,2010MNRAS.406..822M})

Among the mechanisms of AGN feedback that helps shape galaxy evolution, AGN-driven outflows (or galactic winds) stand out as the most dynamically efficient agent (\cite{2014MNRAS.442..440C,2005ARA&A..43..769V,2012MNRAS.426..140D}). In contrast to thermal feedback, which suffers substantial radiative cooling losses, outflows offer a more efficient channel for quenching star formation, regulating SMBH growth, and injecting energy and momentum into the interstellar medium (ISM).
The strength of AGN feedback can be directly assessed by measuring the physical properties of the outflows, such as the outflow mass, mass outflow rate, and kinetic luminosity (\cite{2012ApJ...751..107B}), offering crucial diagnostics for testing theoretical models of AGN feedback. 
Such properties are typically inferred from observable quantities: outflow velocity, column density, and the radial distance from the central AGN.
Unlike velocity, which is attainable from emission or absorption lines,and photionization information, which can be determined from ionic column density measurements, the radial distance of the outflow (i.e. its distance to the SMBH) is an essential parameter for constraining outflow energetics, which remains a challenge for observers.

Absorption lines serve as a fundamental tracer of AGN outflows. The galactocentric distance of absorption outflows is commonly inferred from its relationship with the ionization parameter: $U_H = \frac{Q_H}{4 \pi R^2 n_H c}$. 
In this equation, the hydrogen-ionizing photon rate $Q_H$ can be estimated from the shape of the incident spectral energy distribution (SED), and the ionization parameter $U_H$ can be derived from photoionization equilibrium. While most quantities in this equation can be reliably constrained through spectroscopy and numerical modeling, constraining the density $n_H$ is instrumental to inferring the outflow’s radial distance R.

Over the past decade, numerous efforts have been made to constrain the density of outflowing gas (e.g. \cite{2014MNRAS.444.1893C, 2014ApJ...783...58L, 2011ApJ...728...94L, Arav_2015, Arav_Liu_2018, 2018ApJ...866....7L, 2019MNRAS.483.1808H, Ebrero_2016,2017A&A...607A.100M}). Among the various techniques employed, the most reliable to date has been the analysis of absorption lines from excited ionic states. By measuring the column density ratio between excited and ground (resonant) states, one can infer the electron number density of the absorbing gas.
However, this method requires high signal-to-noise (S/N) ratio spectra and is often hindered by line blending, a prevailing phenomenon in broad and high-velocity outflows, often rendering it difficult to identify clean absorption features for density diagnostics. Nevertheless, such energetic outflows represent the most powerful agents of AGN feedback and are of particular interest for understanding SMBH-galaxy co-evolution, highlighting the necessity to explore alternative approaches to accurately determine outflow distances when conventional methods are inefficient.

In recent years, a viable new method has been developed to estimate outflow distances by monitoring absorption line variability detection rates (\cite{He_2019}). This technique relies on estimating the recombination timescale, i.e. the characteristic response time of absorbing gas to variations in the ionizing continuum. 
\cite{He_2019} directly examines the probability of detecting absorption-line variability as a function of the observational time interval. By capturing the rapid rise in detection probability near the recombination timescale and by fitting the resultant detection rate curve of absorption line variations with a Gaussian cumulative distribution function (CDF), $t_r$ is thus  constrained.
The growing availability of multi-epoch spectroscopic datasets (especially from reverberation mapping campaigns that densely sample individual AGNs), and ongoing refinement of the method and its theoretical underpinnings, this methodology now not only allows for more accurate outflow distance measurements, but also for reconstruction of the velocity profile as a function of radial distance.
\textbf{Although \cite{He_2019} applied this method to a statistical sample of hundreds of quasars, the technique itself is essentially statistical in nature. The physical processes described in the construction of the detection rate curve, such as the delayed response of the absorbing gas to changes in the ionizing continuum and the dispersion of variability timescales arising from the complexity of real astrophysical environments, are the same physical mechanisms underlying outflows in individual sources. Therefore, this method is also applicable to determining outflow radial distances in single objects.}

NGC 5548 is a nearby, prototypical Seyfert 1 galaxy and has long served as a prime target for AGN investigations (\cite{2000A&A...354L..83K, 2013A&A...551L...6K, 2016A&A...592A..27C, 2017MNRAS.464.1771S, 2020ApJ...902...74W, 2022MNRAS.516.4397K, AGN_STORM_2021, Mehdipour_2024}). Owing to its proximity and extensive observations, its variability characteristics, spectral energy distribution (SED), black hole mass, broad-line region (BLR) size and many other properties have been thoroughly analyzed.
Moreover, the accumulation of an exceptionally rich set of multi-epoch spectroscopic observations of this galaxy is spectacular, with historical optical spectra dating back to the early 1970s (\cite{Sergeev_2007}), providing a unique advantage for variability research. In particular, the AGN Space Telescope and Optical Reverberation Mapping (AGN STORM, \cite{AGN_STORM}) campaign in 2014 delivered an intensive, multi-wavelength monitoring program using Swift, the Hubble Space Telescope (HST), and several ground-based optical telescopes. This campaign is featured by its high-cadence, high-quality spectroscopic data ideal for variability-based AGN science.
NGC 5548 does not exhibit classical broad absorption lines (BALs), but its absorption lines allow the traditional method based on excited-to-resonant state column density ratios to deduce outflow radii (\cite{Arav_2015}), facilitating comparison with our results.

In this work, we analyze the variability of the ultraviolet C IV absorption lines in the spectra of NGC 5548, identify the multiple kinematic components of the outflow, and determine their individual distances. The structure of this paper is as follows: Section 2 describes the multi-epoch HST/COS ultraviolet spectroscopic data used in this work. Section 3 details the method and analysis procedure for estimating outflow distances based on absorption line variability. In Section 4, we present the measurements of recombination timescales and the inferred outflow radii. Section 5 discusses the assumptions underlying the method and our comparison with previous works. A summary of our findings is given in Section 6.

\section{Data and Observations}
High-quality multi-epoch ultraviolet (UV) spectroscopic observations are crucial for tracing absorption-line variability and probing the physical conditions of AGN-driven outflows. To this end, we compile multi-epoch archival UV spectra of NGC 5548 and selected the 2014 observations from the AGN STORM project as our primary dataset. Earlier HST/COS observations from 2013 were also included to cover a wider time frame.

As part of the AGN STORM campaign, \cite{AGN_STORM} conducted an intensive 180-day monitoring program of NGC 5548 using the Cosmic Origins Spectrograph (COS) on board the Hubble Space Telescope (HST). The monitoring from February 1 to July 27 in 2014 was on a nearly daily basis with single-orbit exposures, providing a high-cadence sampling of the UV variability in NGC 5548, yielding a high-quality spectroscopic dataset that fully meets the requirements of this study.

Each observation employed both the G130M and G160M gratings, with varying central wavelength and focal-plane position (FP-POS) settings to cover the full UV range of 1153–1796 Å. Each visit included two 200-second exposures with the G130M grating (centered at 1291 Å and 1327 Å) and two 590-second exposures with the G160M grating (centered at 1600 Å and 1623 Å). This setup was optimized to achieve a signal-to-noise ratio (S/N) greater than 100 in each $\sim$500 km/s velocity bin, enabling robust measurements of subtle absorption line variability.

The AGN STORM data used in this work are obtained directly from the Mikulski Archive for Space Telescopes (MAST) \footnote{https://mast.stsci.edu/portal/Mashup/Clients/Mast/Portal.html}. All spectra are reduced using the standard CalCOS pipeline, and the uncertainty of relative flux calibration is  within 2\%.

Subsequent works (\cite{2016ApJ...824...11G, 2019MNRAS.486.5362G,2017ApJ...837..131P}) reveal an anomalous period during the AGN STORM campaign (JD 2456766.1 to JD 2456829.8, or roughly 60 days), during which the broad emission-line variability in the UV–optical band became decoupled from the far-UV continuum variations. This behavior has been interpreted as arising from either changes in the unobservable ionizing continuum that affect its strength, spectral shape, or temporal structure, or from a transient obscurer located between the ionizing source and the BLR, which selectively attenuates the ionizing flux reaching the BLR without affecting the continuum observed along our line of sight.

We restrict our analysis to the data obtained prior to the onset of the anomalous period (i.e., before JD 2456766.1) due to the following reason. Our work is based on the assumption that variability of absorption lines is fundamentally driven by changes in the incident ionizing continuum. While the specific mechanisms behind the continuum variability (e.g., intrinsic changes or obscuration) are not of direct concern here, the outflowing gas may not respond instantaneously due to recombination timescales. 
Therefore, the corresponding data were excluded from the final analysis to avoid potential contamination (for reference,  the specific time range and the corresponding continuum light curve can be seen in the unshaded portion preceding the gray-shaded anomalous period in Figure 8 of \cite{2019ApJ...881..153K}).

To further expand our time frame, we also incorporate earlier HST/COS UV spectra obtained in 2013 (\cite{Kaastra_2014_ngc5548_2013_campaign,Mehdipour_2015_ngc5548_2013_campaign}), consisting of five COS visits conducted between 2013 June 22 and August 1. Each visit comprised two HST orbits, using both G130M and G160M gratings with multiple central wavelength and FP-POS settings. The resultant spectra cover the 1132–1801 Å range at a resolving power of $\sim 15000$.

In total, our final data set comprises 76 UV spectroscopic observations covering multiple 2013 and 2014 epochs.

\section{Theoretical Framework and Implementation}
\label{sec:3}
To constrain the radial distance of AGN outflows using variability in absorption lines, we adopt a method based on two fundamental relations (\cite{Arav_2012}): 

\begin{equation}
    t_r = \left[-f\alpha_{\mathrm{CIV}}n_e\left(\frac{n_{\mathrm{CV}}}{n_{\mathrm{CIV}}} - \frac{\alpha_{\mathrm{CIII}}}{\alpha_{\mathrm{CIV}}}\right)\right]^{-1},  \label{eq:1} 
\end{equation}

\begin{equation}
    U_H = \frac{Q_H}{4 \pi R^2 n_H c}, \label{eq:2}
\end{equation}
where $t_r$ is the recombination timescale of the absorbing gas, and $f$ represents the fractional change in the ionizing continuum intensity between two observational epochs, for which a typical value of 0.1 is adopted in this work (\cite{He_2022}).
The recombination rate coefficient $\alpha_i$ corresponds to recombination from ionization stage $i+1$ to $i$. These coefficients are taken from the CHIANTI atomic database version 8.0 \citep{CHIANTI} at a temperature of $2 \times 10^4,\mathrm{K}$, yielding $\alpha_{\mathrm{C\,III}} = 2.1 \times 10^{-11}\,\mathrm{cm^3\,s^{-1}}$ and $\alpha_{\mathrm{C\,IV}} = 5.3 \times 10^{-12}\,\mathrm{cm^3\,s^{-1}}$. 
The ion density ratio $n_{\mathrm{CV}}/n_{\mathrm{CIV}}$ is derived from photoionization equilibrium models computed with the CLOUDY code (\cite{Cloudy2023}), based on the best-fit values of the ionization parameter $U_H$ and hydrogen column density $N_H$ (see Appendix~\ref{appendix:cloudy}).
\textbf{In addition, we would like to emphasize that the definition of $t_r$ and equation~(\ref{eq:1}) originates from the ionization–recombination equilibrium of the gas. Therefore, the timescale $t_r$ inherently incorporates both the ionization and recombination processes, and it is well defined when the change in ionizing flux is relatively small.}

Equation (\ref{eq:2}) gives the classical definition of the ionization parameter $U_H$, where $Q_{\rm H}$ is the rate of hydrogen-ionizing photon emission (for photon energies $>13.6$ eV) from the central engine, $c$ is the speed of light, $R$ is the radial distance from the ionizing source to the absorber, and $n_H$ is the total hydrogen number density.
Given that C IV is a representative high-ionization species observed in AGN spectra, we adopt the approximation $n_H \approx 0.83\,n_e$, assuming a fully ionized gas where hydrogen and helium dominate the composition and the contribution from metals is negligible.

Consequentially, variations in absorption lines provide a means to estimate the absorber’s radial distance through pinning down the recombination timescale $t_r$.

\subsection{Determining the Recombination Timescale via Detection Rate of Absorption Line Variability}
The recombination timescale $t_r$ characterizes the delay with which absorption lines respond to variations in the incident ionizing continuum. For a given observational time interval $\Delta T$, the gas response can be modeled by a step-function response probability (\cite{He_2019}):
\begin{equation}
p(t_r, \Delta T) = 
\begin{cases} 
1 & \text{if } t_r \leq \Delta T, \\
0 & \text{if } t_r > \Delta T.
\end{cases}
\end{equation}
That is, variability in the absorption lines becomes observable only when the observational interval $\Delta T$ exceeds the recombination timescale $t_r$, allowing the gas to fully respond to changes in the ionizing continuum.

Owing to intrinsic fluctuations in ionization state and velocity, the recombination timescale of an outflowing gas component is appropriately characterized by a Gaussian distribution:
\begin{equation}
f(t_r) = \frac{1}{\sqrt{2\pi\sigma^2}} \exp\left(-\frac{(t_r - \mu)^2}{2\sigma^2}\right),
\end{equation}
where $\mu$ is the characteristic timescale and $\sigma$ denotes the dispersion. Hence, the probability of detecting absorption-line variability within a time interval $\Delta T$ is given by:
\begin{equation}
F(\Delta T) = \int_0^{+\infty} p(t_r, \Delta T)f(t_r) \, dt_r = \int_0^{\Delta T} f(t_r) \, dt_r.
\end{equation}
Intuitively, this formulation renders an observational trend: when $\Delta T < t_r$, the integral $F(\Delta T)$ accumulates only a small portion of the probability density, resulting in a low likelihood of detecting variability; but for $\Delta T > t_r$, a larger fraction of the distribution contributes to the integral, and the detection probability increases accordingly. 
In observations, variability in absorption lines is scarcely seen when $\Delta T$ is shorter than the recombination timescale, as the gas has not yet fully responded to continuum changes. Once $\Delta T$ surpasses $t_r$, the detection likelihood rises steeply.

This characteristic response lays the foundation for our detection rate curve method, in which we statistically quantify the detection rate of absorption line variability as a function of $\Delta T$ to constrain the recombination timescale $t_r$ (\cite{He_2019}).


\subsection{Detection of Absorption Line Variability}
To measure variability in absorption troughs manifesting the AGN outflow, we utilize a direct spectral comparison method rather than relying on traditional equivalent width (EW) measurements derived from continuum fitting.

For each pair of spectroscopic observations, the spectrum with the higher S/N is taken as the reference spectrum, which is depicted by a double power-law function, with additional Gaussian components added near the C IV emission line to account for line variability. 
In particular, the emission profile of C IV in NGC 5548 features both a spike and an extended base, motivating the use of two Gaussian components. 
To prevent the narrow Gaussian from falling into the absorption trough, we set its central wavelength within the range 1549.7-1550.4 \AA.
Furthermore, the impact of absorption features on the overall fit is mitigated by down-weighting these spectral regions by applying a multiplicative factor of 0.3 to the inverse-variance weights.

The double power-law function used to depict the reference spectrum is given by:
\begin{align}
S_\lambda = A[1]\left(\frac{\lambda}{2000\,\text{\AA}}\right)^{A[2]} + A[3]\left(\frac{\lambda}{2000\,\text{\AA}}\right)^{A[4]},
\label{eq:double-power-law}
\end{align}
where $A[1]$ to $A[4]$ are free parameters.
Following this step of spectral matching, we examine the blue side of the C IV absorption feature on a pixel-by-pixel basis to identify flux variations. The detection criteria listed below are defined on the basis of variability behavior observed in the data:

Candidate variable regions are initially identified as contiguous segments of at least three pixels exhibiting flux changes in the same direction. 
A region is retained if a flux deviation exceeds $3\sigma$ in at least one pixel, and exceeds $1\sigma$ in at least one of its neighboring pixels. Regions separated by fewer than four pixels are merged, while regions shorter than five pixels are discarded. The integrated significance of each region is then quantified using:
$$
N_\sigma = \frac{|\sum \Delta \mathrm{flux}|}{\sqrt{\sum \sigma^2}}.
$$
Only those regions with $N_\sigma > 3$ are considered to show statistically significant variability. Examples of detected absorption line variability are presented in Figures~\ref{fig:var_examples_short} and \ref{fig:var_examples_long}.

\subsection{Determining the Recombination Timescale $t_r$}
Based on the principles outlined above, we construct the detection rate curve of absorption line variations, $\Delta T$--$F(\Delta T)$, by computing the fraction of spectral pairs exhibiting absorption-line variability as a function of the observational time interval $\Delta T$. By fitting this curve to the Gaussian CDF, we derive the characteristic recombination timescale $t_r$ for the outflowing gas.

When multiple kinematic components exist in an absorption trough, the $t_r$ distribution becomes a combination of multiple Gaussians, and the detection rate curve can be modeled as the sum of a number of Gaussian CDFs, allowing for measuring recombination timescales associated with individual components.


There exist 2850 different pairs among the 76 spectra of NGC 5548 under consideration. 
To ensure sufficient variation in the ionizing continuum while minimizing the loss of spectral pairs at short time intervals, we apply a selection criterion of $|\Delta L/L|_{1500\,\text{\AA}} \geq 20\%$, yielding a final sample of 1530 pairs. 
Here, the relative continuum variation is defined as
\begin{equation}
    |\Delta L / L| = \left| \frac{2(L_2 - L_1)}{L_1 + L_2} \right|,
\end{equation}
where $L_1$ and $L_2$ are the continuum fluxes measured at the first and second epoch, respectively.
For each spectral pair and absorption region, we calculate the total variability significance $N_\sigma$ in the C\,IV absorption troughs. Pairs with $N_\sigma > 3$ are regarded to exhibit significant absorption-line variability.

We compute the detection rates of absorption line variations  in those bins containing equal amounts of spectral pairs, typically we divide the sample into 30 bins ($\sim$52 pairs per bin). For Trough F, 35 bins ($\sim$44 pairs per bin) are used. The detection rate in each bin, $F_i = k_i/N_i$ (with $k_i$ the number of observations that show variability and $N_i$ the number of spectral pairs contained in the same bins), and its associated binomial uncertainty are calculated as
$$
\sigma_{F_i} = \sqrt{\frac{F_i (1 - F_i)}{N_i}}
$$
The representative time interval for each bin, denoted as $\Delta T_i$, is defined as the arithmetic mean of all time intervals within that bin. 
The detection rate curve $F(\log_{10} \Delta T)$ is then constructed by plotting $F_i$ against $\log_{10} \Delta T_i$.

To characterize the rapid increase in detection rate near $t_r$, we fit the curve to the Gaussian CDF
$$
F(x) = \frac{A}{2}\left[1 + \mathrm{erf}\left(\frac{x-\mu}{\sigma\sqrt{2}}\right)\right],
$$
where $\mu$ corresponds to the logarithmic recombination timescale, $\log_{10} (t_r)$.

Due to the limited wavelength coverage of the C\,IV absorption profile in NGC 5548 (only about four times the separation of the C\,IV doublet), the red component of a high-velocity system may coincide with the blue component of a lower-velocity system (see Figure~\ref{fig:1}), resulting in multiple kinematic components contributing to a single absorption trough. When these components exhibit significantly different recombination timescales, the resulting detection rate curve may show a ``multi-step'' profile. In this work, two absorption troughs (E and F) display evident double-step features, for which we model the curve using a two-component Gaussian CDF:
$$
F(x) = \frac{A_1}{2}\left[1 + \mathrm{erf}\left(\frac{x-\mu_1}{\sigma_1\sqrt{2}}\right)\right] + \frac{A_2}{2}\left[1 + \mathrm{erf}\left(\frac{x-\mu_2}{\sigma_2\sqrt{2}}\right)\right],
$$
which allows for independent characterization of each component’s response and facilitates estimation of their respective recombination timescales.

The detection rate curves of absorption line variations for all measured absorption troughs are presented in Figures~\ref{fig:troughs_unconstrained}, \ref{fig:trough8_9_detection}, and \ref{fig:trough5_6_detection}. 
For troughs that exhibit sufficiently broad detection rate distributions and can be successfully fit to a Gaussian CDF, the best-fit models are shown as blue curves.
In these cases, a supplementary panel is given below each main plot to display the corresponding Gaussian distribution of the recombination timescale $t_r$ for the associated outflowing component.

\subsection{Outflow Velocity Components and Absorption Trough Segmentation}

UV outflows in NGC 5548 are found to show six velocity components over the past two decades (\cite{Crenshaw_2003}), which are also associated with warm absorber (WA) components detected in the X-ray. In this work, we adopt the same strategy, but recompute the velocities using an updated redshift of $z = 0.017175 \pm 0.000023$ (\cite{1991rc3..book.....D}), and make minor adjustments to the velocity intervals based on the actual distribution of absorption troughs in our spectra. The revised velocity intervals are listed in Table~\ref{table:D22}.

\begin{figure}
    \centering
    \includegraphics[width=0.75\linewidth]{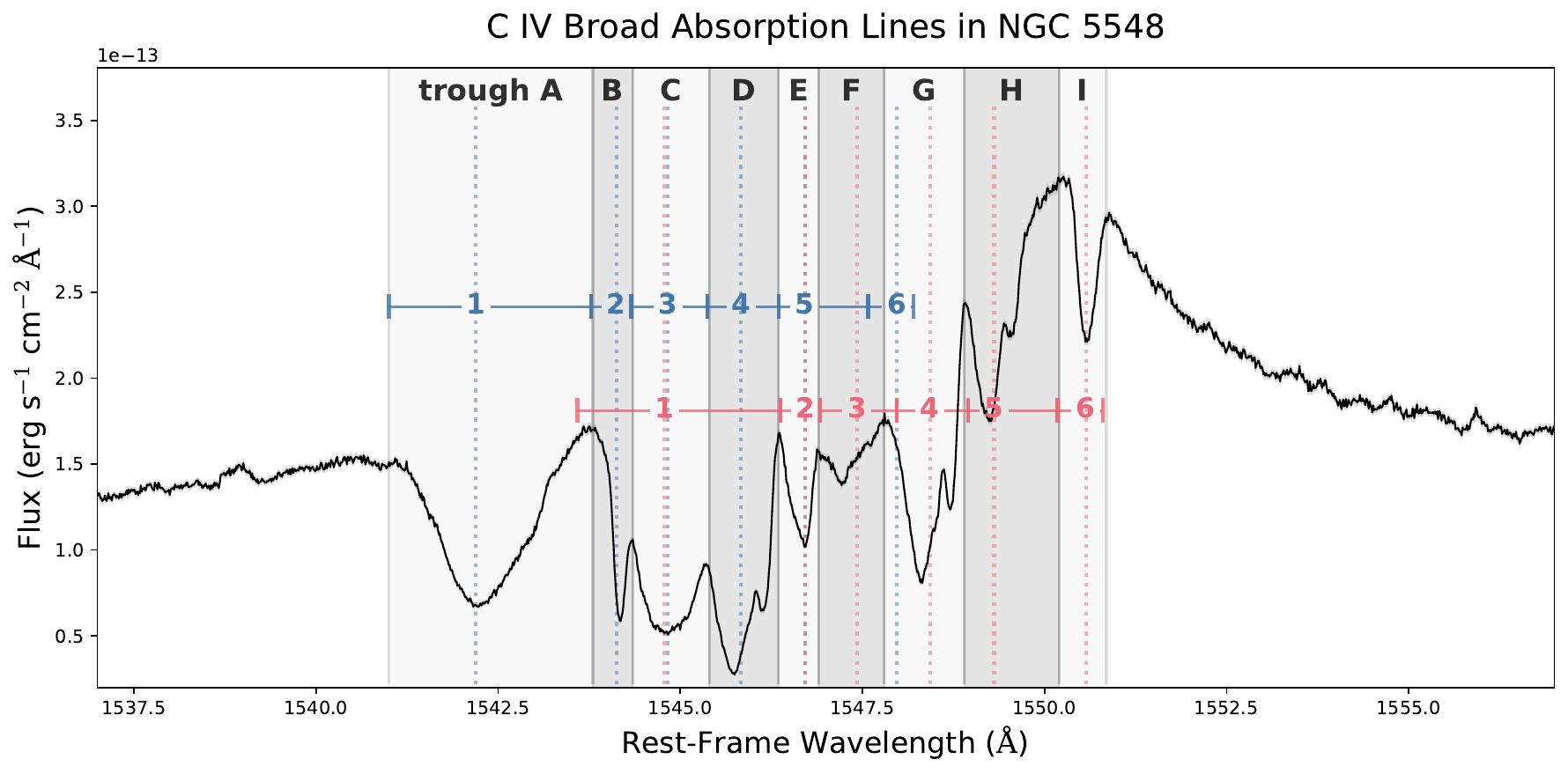}
    \caption{\textbf{Broad C IV absorption lines in NGC 5548.} 
    The black curve shows is geometric mean composite spectrum of all observed UV spectra. Red and blue horizontal bars and overlaid numbers indicate the velocity ranges of the six ultraviolet outflow components (\cite{Crenshaw_2003}), with blue and red bars corresponding to the blue and red members of the C IV doublet, respectively. Vertical dashed lines mark the central wavelengths of each component. Alternating shaded vertical bands represent the absorption troughs analyzed in this study, with their numerical labels shown at the top in boldface.}
    \label{fig:1}
\end{figure}

Figure~\ref{fig:1} shows a composite spectrum generated from the geometric mean of all 76 available spectra. Colored horizontal lines and numeric labels mark the locations of the six velocity components. Red and blue markers represent the red and blue members of the C\,IV doublet within each component, respectively. As illustrated, the doublets from different components overlap substantially. 

By mapping the C\,IV doublet transitions of each velocity component into wavelength space, we assess the overlapping contributions of multiple components to each trough. The resulting component-trough associations, ordered by increasing wavelength, are listed as follows (also see Table~\ref{tab:overlap_Comp}):

\begin{itemize}
    \item	Trough A: blue member of velocity component 1;
    \item	Trough B: red member of velocity component 1 and blue member of velocity component 2;
    \item	Trough C: red member of velocity component 1 and blue member of velocity component 3;
    \item	Trough D: red member of velocity component 1 and blue member of velocity component 4;
    \item	Trough E: red member of velocity component 2 and blue member of velocity component 5;
    \item	Trough F: red member of velocity component 3, blue member of velocity component 5, and minor contribution from blue member of velocity component 6;
    \item	Trough G: red member of velocity component 4, minor contribution from red member of velocity component 3 and blue member of velocity component 6;
    \item	Trough H: red member of velocity component 5;
    \item	Trough I: red member of velocity component 6.
\end{itemize}

\begin{deluxetable}{c|ccc}
\tablecaption{Overlapping Contributions of Multiple Components to Absorption Troughs\label{tab:overlap_Comp}}
\tablehead{
\colhead{Absorption Trough} & \multicolumn{3}{c}{Contributions of Components}
}
\startdata
A & Blue of Comp. 1     & \nodata             & \nodata \\
B & Red of Comp. 1      & Blue of Comp. 2     & \nodata \\
C & Red of Comp. 1      & Blue of Comp. 3     & \nodata \\
D & Red of Comp. 1      & Blue of Comp. 4     & \nodata \\
E & Red of Comp. 2      & Blue of Comp. 5     & \nodata \\
F & Red of Comp. 3      & Blue of Comp. 5     & Blue of Comp. 6 (minor) \\
G & Red of Comp. 4      & Blue of Comp. 6     & Red of Comp. 3 (minor)  \\
H & Red of Comp. 5      & \nodata             & \nodata \\
I & Red of Comp. 6      & \nodata             & \nodata \\
\enddata
\end{deluxetable}

\section{Results}

The recombination timescales $t_r$ of the outflow components derived from the detection rate curves for each absorption trough, combined with the ionic column densities of C IV and Si IV derived from trough fits and the ionization parameters obtained via \texttt{CLOUDY} modeling (see Appendix~\ref{appendix:cloudy}), are prerequisites for estimating the radial distance of the outflowing components from the central engine. Our results are summarized in Table~\ref{table:D22}.

\subsection{Absorption Troughs A to D}

Troughs A, B, C, and D all receive contribution from velocity component 1. As shown in Figure~\ref{fig:troughs_unconstrained}, the detection rate of absorption line variations in trough A (the blue member of velocity component 1 only) is constantly above 80\%. According to the theoretical framework outlined in Section~\ref{sec:3}, a monotonically increasing detection rate emerges only when $t_r$ is longer the observational time intervals. The absence of such a trend therefore implies the opposite, i.e., $t_r < 2.51$ days (under the $|\Delta L/L|_{1500,\text{\AA}} \geq 20\%$ selection criterion). 

Troughs B to D affected by velocity component 1 show similarly high detection rates: always over 88\% for trough C, and all but one point (corresponding to the longest time interval) for trough D over 85\%, with a minimum of 73\%. Trough B shows a marginally increasing trend, with all points above 74\%.

The detection rate curve of trough A cannot be fitted with a Gaussian CDF. For troughs B, C and D, the contribution of velocity component 1 overwhelms that of other kinematic components, preventing us from deducing reliable recombination timescales.

For velocity component 1, we constrain its radial distance to be $R < 3.80$~pc, based on $t_r < 2.51$ days and our \texttt{CLOUDY} modeling results ($U = -1.81^{+0.13}_{-0.12}$, $n_{\mathrm{C\,V}}/n_{\mathrm{C\,IV}} = 0.579^{+0.125}_{-0.046}$).

\subsection{Absorption Troughs H and I}

Troughs H and I are each contributed by a single velocity component, and $t_r$ is thus given by fits to a single Gaussian CDF. We exclude data points with time intervals longer than 100 days, as we find that nearly all detection rate curves drop therein for unknown reasons \citep{He_2019,qingyuan2021}. The remaining data points in both troughs fit the model well, as illustrated in Figure~\ref{fig:trough8_9_detection}.

The detection rate curve for Trough H (corresponding to component 5) exhibits a well-defined steadily increasing trend, rising from 41\% at $\log \Delta T = 0.685$ to 75\% at $\log \Delta T = 1.5$.
For velocity component 5, we derive a recombination timescale of $t_r = 4.41 \pm 0.67$ days. Along with $U = -1.00^{+0.03}_{-0.04}$ and $n_{\mathrm{C\,V}}/n_{\mathrm{C\,IV}} = 4.934^{+0.943}_{-0.133}$ from \texttt{CLOUDY} modeling), we find a radial distance of $R = 0.97^{+0.26}_{-0.19}$ pc, a result of Equations~\ref{eq:1} and \ref{eq:2}, together with a Monte Carlo simulation that renders the asymmetric errorbar.

Trough I exhibits an overall lower detection rate, which may be attributed to its proximity to the peak of the C IV emission line. As a relatively weak and isolated feature, it is spatially separated from the main absorbers. The derived ionic column density carries large uncertainties, rendering the resultant ionization parameter no better than a lower limit.

For velocity component 6, the recombination timescale is found to be $t_r = 4.83 \pm 1.28$ days. The corresponding \texttt{CLOUDY} modeling yields $U = -0.56^{+\infty}_{-0.71}$ and $n_{\mathrm{C\,V}}/n_{\mathrm{C\,IV}} = 15.977^{+1.485}_{-0.270}$, resulting in an estimated radial distance of $R = 2.81^{+2.92}_{-2.81}$~pc.

\subsection{Absorption Troughs E and F}

Troughs E and F uniquely exhibit distinct ``multi-step'' features in their detection rate curves. These pronounced profiles allow us to effectively disentangle and constrain the recombination timescales of the velocity components that contribute to it.
As shown in Figure~\ref{fig:trough5_6_detection}, both curves begin with a gradual increase in detection rate from relatively low values, then an abrupt rise occurs, rapidly increasing from 60--70\% to nearly 100\%. 

To model these segmented responses, we employ a two-Gaussian CDF. The results show that the shorter recombination timescales fall within the range of a few days, while the second, steeper rise corresponds to much longer timescales, on the order of $\sim$30 days.

From visual inspection of the spectra and the overlapping positions of C IV doublet components, we find that trough E is influenced by the red member of velocity component 2 and the blue member of component 5, while trough F by the red member of component 3 and the blue members of components 5 and 6. Based on the timescales derived for components 5 and 6 from troughs H and I, which are both on the order of days, we infer that the shorter $t_r$ components in the ``multi-step'' profiles of troughs E and F originate from components 5 and 6, thus the longer-timescale components (on the order of ~30 days) have to be attributed to components 2 and 3.

We thus conclude that velocity component 2 has a recombination timescale of $t_r = 36.06 \pm 2.23$ days, with \texttt{CLOUDY} modeling that gives $U = -2.63^{+0.30}_{-0.34}$ and $n_{\mathrm{C\,V}}/n_{\mathrm{C\,IV}} = 0.101^{+0.000}_{-0.000}$. These parameters lead to a radial distance of $R = 43.83^{+15.73}_{-12.27}$~pc. 

Similarly, for velocity component 3, we obtain $t_r = 37.15 \pm 1.50$ days, $U = -2.56^{+0.10}_{-0.11}$, and $n_{\mathrm{C\,V}}/n_{\mathrm{C\,IV}} = 0.099^{+0.003}_{-0.001}$, corresponding to a radial distance of $R = 36.63^{+4.03}_{-3.88}$~pc.

\subsection{Absorption Trough G}

\begin{figure}[htbp]
    \centering
    \begin{subfigure}[b]{0.48\textwidth}
        \centering
        \includegraphics[width=\textwidth]{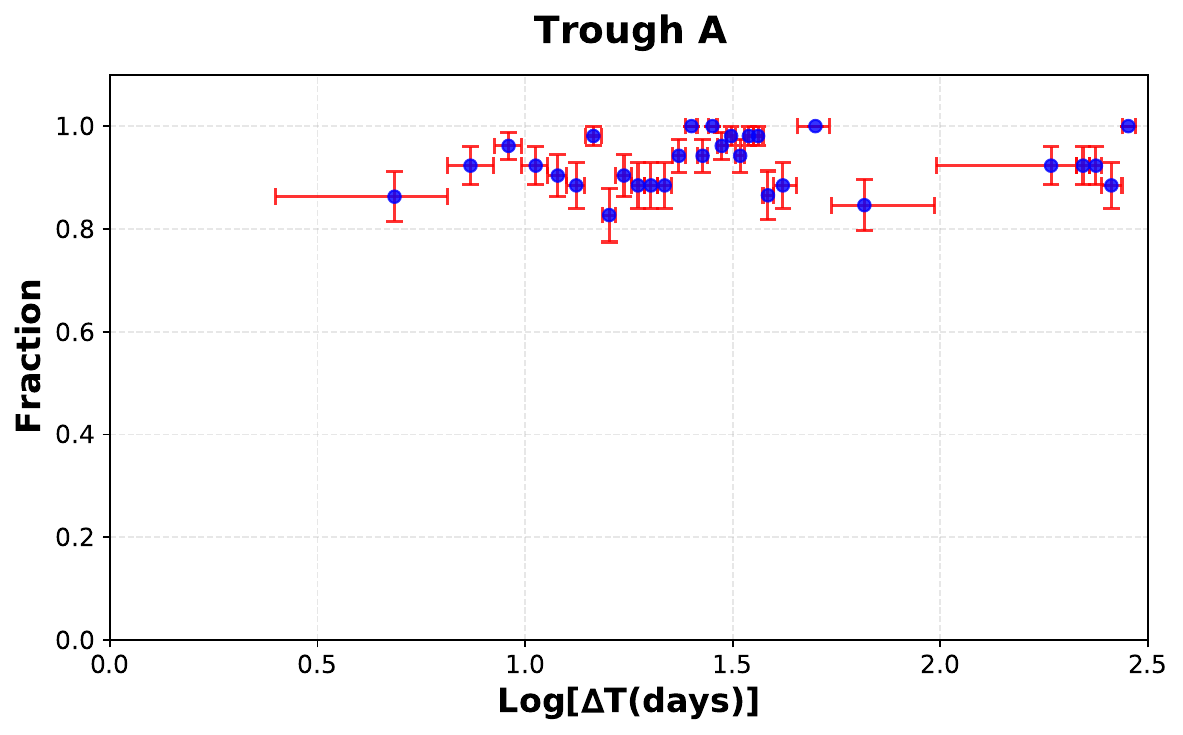}
    \end{subfigure}
    \hfill
    \begin{subfigure}[b]{0.48\textwidth}
        \centering
        \includegraphics[width=\textwidth]{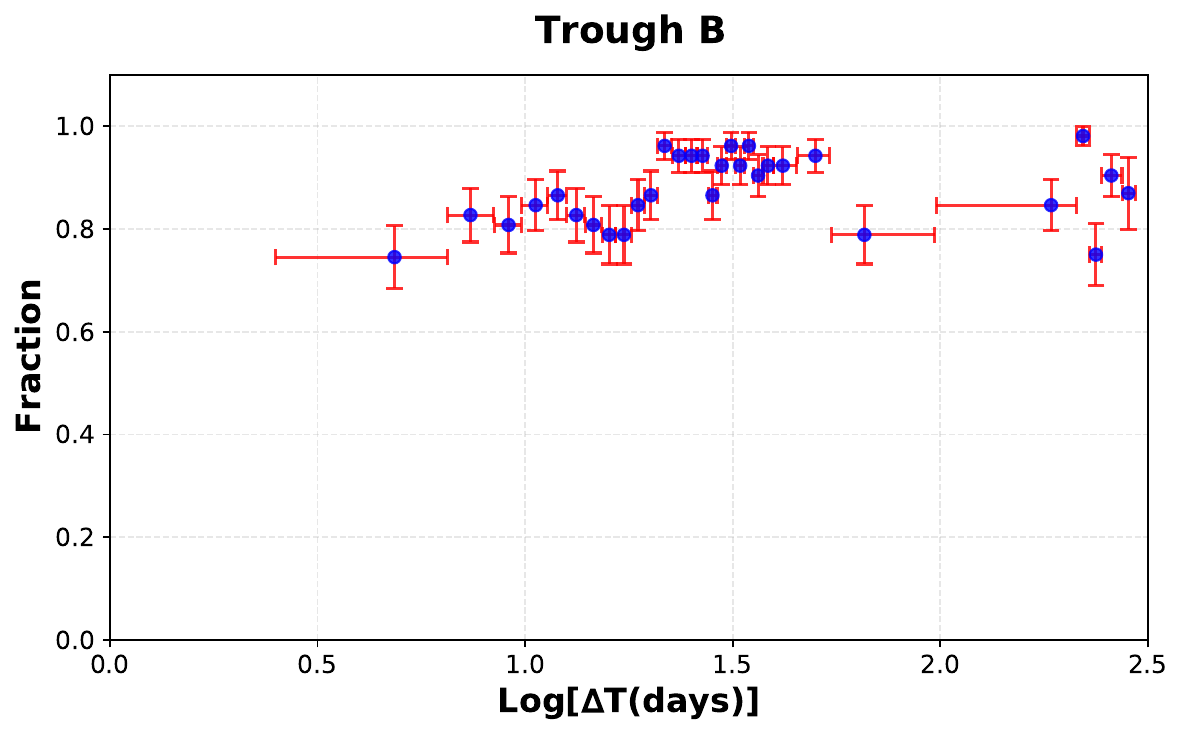}
    \end{subfigure}

    \begin{subfigure}[b]{0.48\textwidth}
        \centering
        \includegraphics[width=\textwidth]{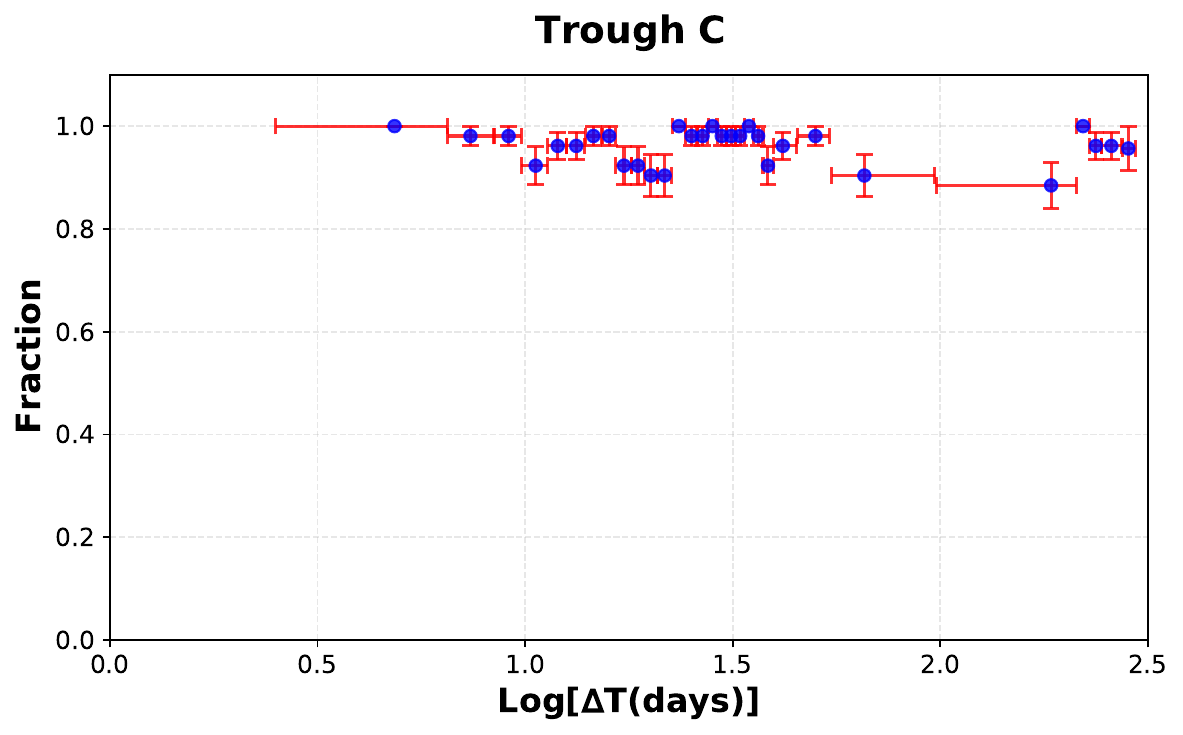}
    \end{subfigure}
    \hfill
    \begin{subfigure}[b]{0.48\textwidth}
        \centering
        \includegraphics[width=\textwidth]{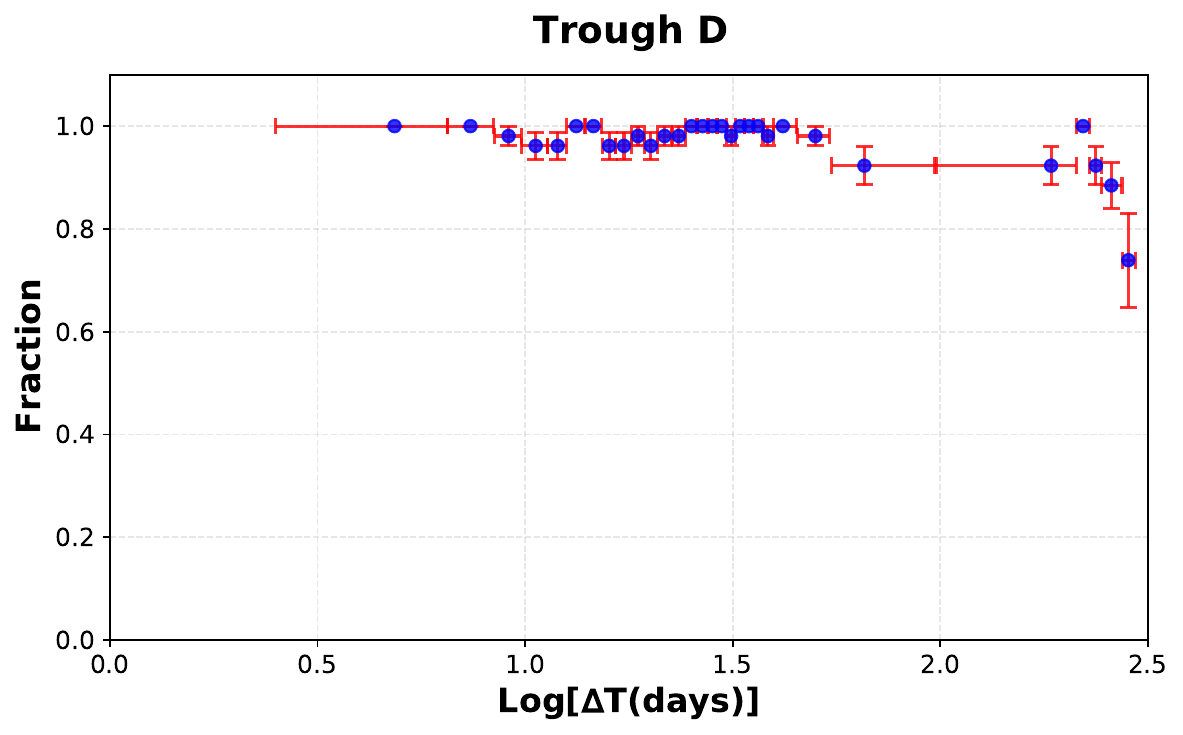}
    \end{subfigure}

    \begin{subfigure}[b]{0.48\textwidth}
        \includegraphics[width=\textwidth]{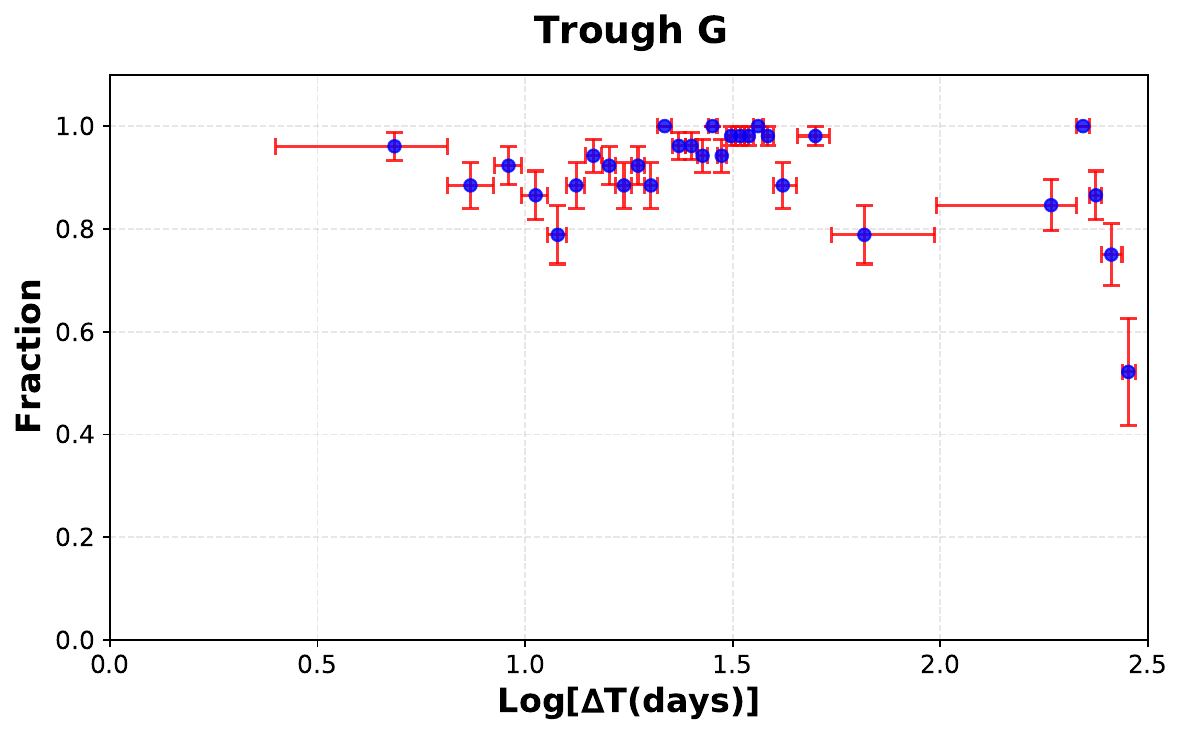}
    \end{subfigure}
    \caption{
    Detection rate distributions for four absorption troughs (Troughs A–D and G), illustrating cases where the recombination timescale $t_r$ cannot be reliably constrained using the detection probability curve method. Each panel shows the fraction of variability detections in absorption troughs ($N_\sigma > 3$) on the vertical axis, as a function of the logarithmic time interval $\log \Delta T$ (in days). The horizontal position and error bar indicate the mean and full range of $\Delta T$ within each bin, while the vertical error bar represents the binomial uncertainty in the detection rate of absorption line variations.
    Trough A corresponds to velocity component 1, whose recombination timescale is significantly shorter than the typical spectroscopic sampling interval of $t_r < 2.51$ days (under the $|\Delta L/L|_{1500,\text{\AA}} \geq 20\%$ selection criterion), resulting in uniformly high detection probabilities. 
    Troughs B to D are significantly contaminated by component 1, leading to similarly high detection rates that obscure any sharp rise near $t_r$; although Trough B shows a single bin with a modestly lower detection rate, the deviation alone lacks statistical significance and is insufficient to constrain $t_r$.
    Trough G is influenced by multiple blended components and lacks a distinct multi-step profile. However, based on known properties of contributing components and its high detection rate, we infer $t_r < 2.51$ days.
    }
    \label{fig:troughs_unconstrained}
\end{figure}

\begin{figure}[htbp]
    \centering
    \begin{subfigure}[t]{0.48\textwidth}
        \includegraphics[width=\textwidth]{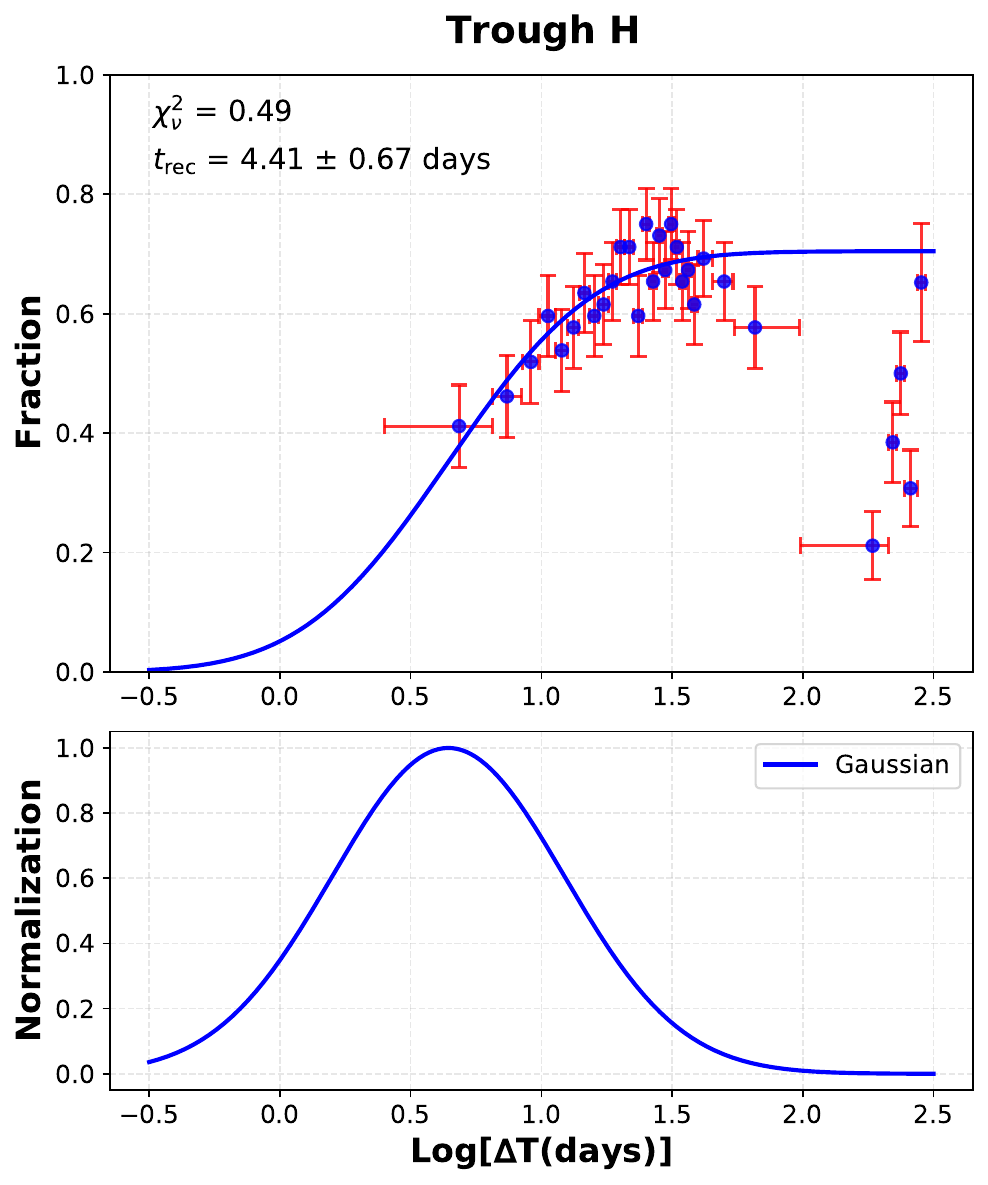}
    \end{subfigure}
    \hfill
    \begin{subfigure}[t]{0.48\textwidth}
        \includegraphics[width=\textwidth]{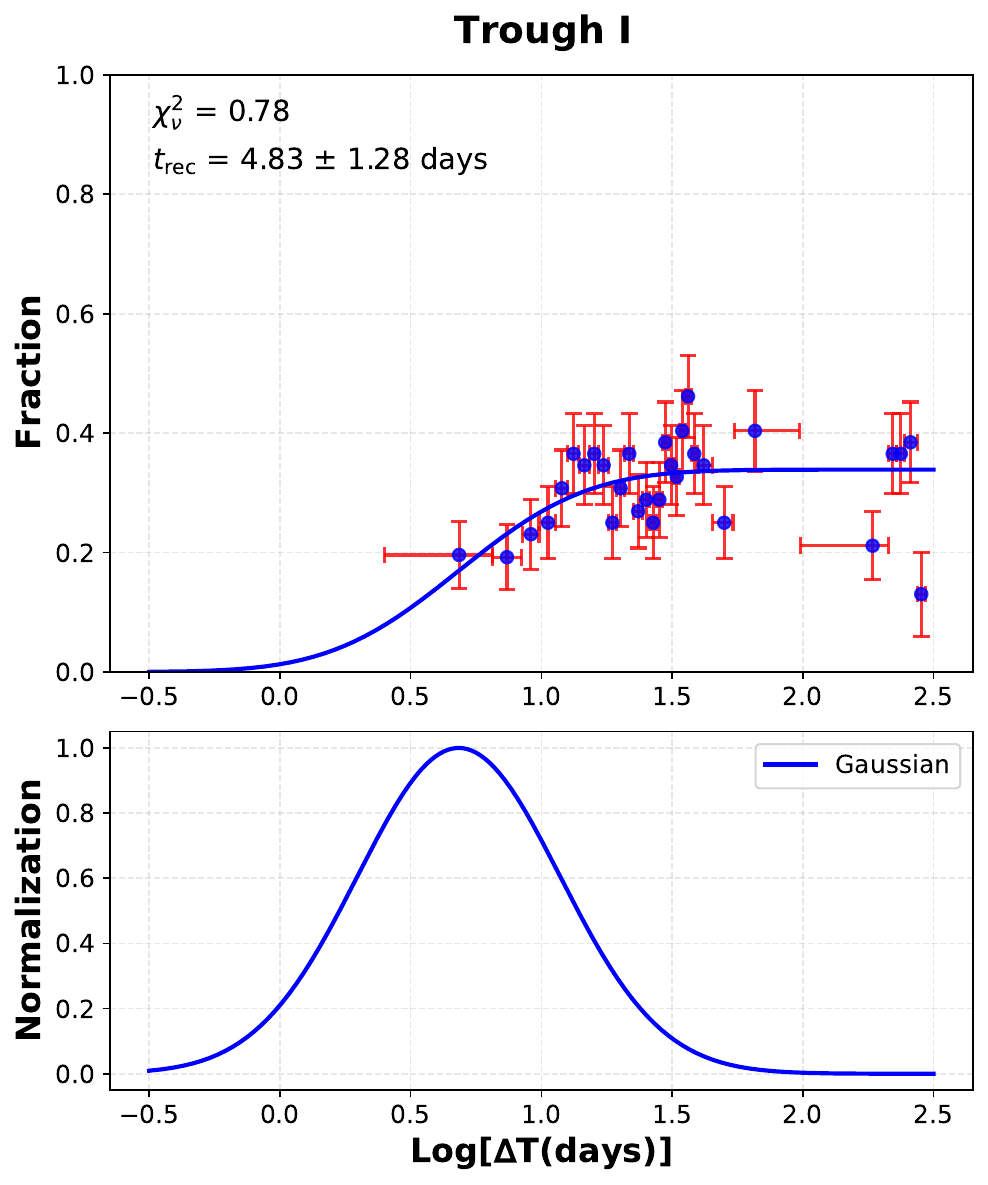}
    \end{subfigure}
    \caption{\textbf{Detection rate curves for Troughs H and I, each predominantly influenced by a single velocity component (components 5 and 6, respectively)}. Both curves are well described by a single Gaussian CDF, enabling reliable estimation of the recombination timescale $t_r$.  
    The upper panel in each subfigure shows the variability detection fraction ($N_\sigma > 3$) as a function of the logarithmic time interval $\log \Delta T$ (in days), consistent with the plotting conventions used in the previous figure. Each data point represents the arithmetic mean of $\Delta T$ values (in linear space) within the corresponding bin, and the horizontal error bars indicate the full range of $\Delta T$. When displayed on a logarithmic scale, the transformation naturally results in visually asymmetric error bars. The blue curve denotes the best-fit Gaussian CDF model, while the lower panel displays the corresponding Gaussian probability density function.  
    For Trough H (component 5), the detection rate fit yields $t_r = 4.41 \pm 0.67$ days. For Trough I (component 6), the fitted $t_r = 4.83 \pm 1.28$ days. 
    }
    \label{fig:trough8_9_detection}
\end{figure}

\begin{figure}[htbp]
    \centering
    \begin{subfigure}[t]{0.48\textwidth}
        \includegraphics[width=\textwidth]{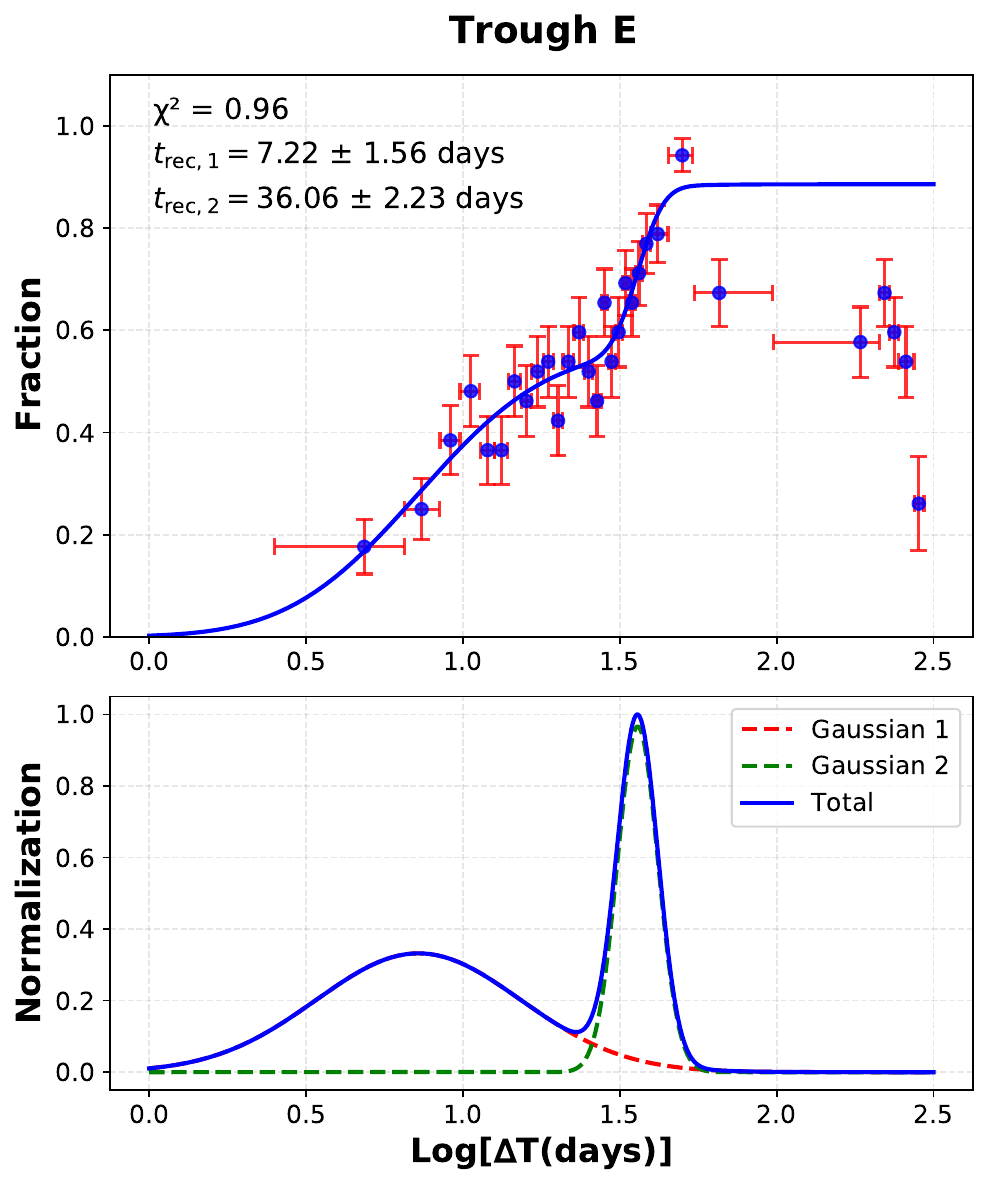}
    \end{subfigure}
    \hfill
    \begin{subfigure}[t]{0.48\textwidth}
        \includegraphics[width=\textwidth]{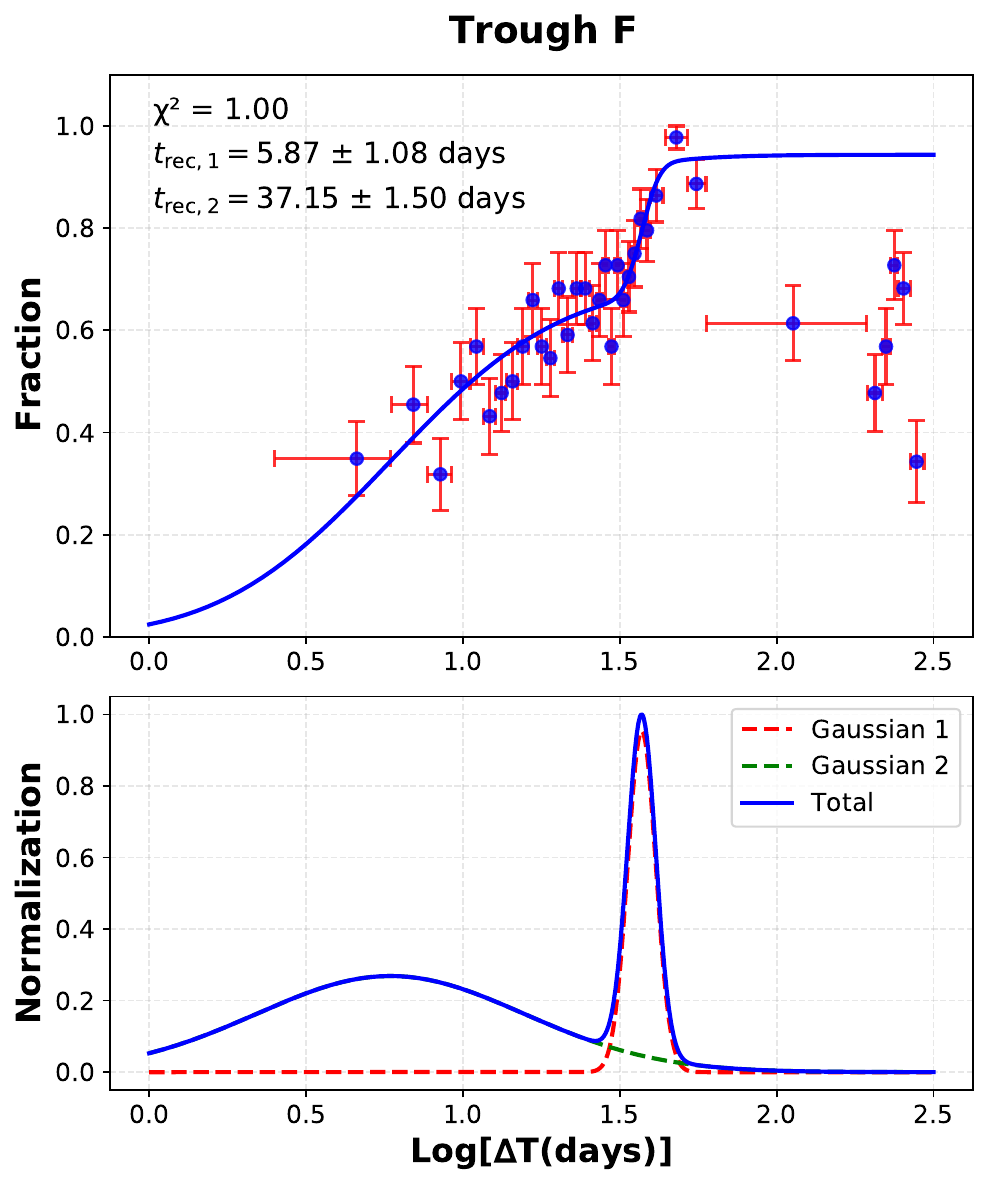}
    \end{subfigure}
    \caption{\textbf{Detection rate curves with double-step profiles for Troughs E and F.} Both curves are well described by a double Gaussian CDF.
    The upper panel in each subfigure shows the variability detection fraction ($N_\sigma > 3$) as a function of $\log \Delta T$ (days), following the conventions of the previous figure. Points denote the arithmetic mean of $\Delta T$ (in linear space) within each bin; horizontal error bars indicate the full range. Logarithmic scaling results in visually asymmetric bars. The blue curve shows the best-fit double Gaussian CDF model.
    The lower panel displays the corresponding Gaussian probability density functions, where the blue solid line indicates the total profile, and the red and green dashed lines represent the two individual Gaussian components. Both components successfully reproduce the two distinct increases in detection probability that define the double-step profiles.
    Trough E is influenced by components 5 and 2, and Trough F by components 6 and 3. Since the $t_r$ values of components 5 and 6 are known to be no more than a few days, the longer timescale in each case can naturally be attributed to component 2 or 3.
    }
    \label{fig:trough5_6_detection}
\end{figure}

As shown in the last panel of Figure~\ref{fig:troughs_unconstrained}, the detection rate curve for trough G is consistently above 78\%, if the decline at longer time intervals is excluded from consideration. Spectral inspection indicates that this trough is primarily influenced by the blue member of velocity component 4, along with minor contribution from the blue member of component 6 and minimal contribution from the red member of component 3.

As component 3 has a relatively long recombination timescale of $t_r = 37.15 \pm 1.50$ days and minimal contribution to the trough, its impact on the detection profile is insignificant. Component 6 has a relatively short recombination timescale ($t_r = 4.83 \pm 1.28$ days), but overall exhibits low detection rates. As trough I lies near the C IV emission peak, the impact of the emission line on its detectability is uncertain. Nonetheless, the consistently high detection rate observed in this trough resembles those of troughs A to D, which are dominated by velocity component 1 with $t_r < 2.51$ days.

Based on this analogy, we infer that velocity component 4 also likely has a recombination timescale shorter than the typical observational interval, i.e., $t_r < 2.51$ days (under the $|\Delta L/L|_{1500,\text{\AA}} \geq 20\%$ selection criterion).
Using the \texttt{CLOUDY} model with $U = -1.53^{+0.02}_{-0.02}$ and  $n_{\mathrm{C\,V}}/n_{\mathrm{C\,IV}} = 1.396^{+0.049}_{-0.022}$, we constrain the radial distance to be $R < 2.29$~pc for this component.

\begin{deluxetable*}{cccccccc}
\tablecaption{Physical Parameters of the C\,\textsc{iv} Outflow Components in NGC 5548 \label{table:D22}}
\tablehead{
\colhead{Component} & \colhead{$v_c$\tablenotemark{a}} & \colhead{Velocity Range} & \colhead{log $U_{\mathrm{H}}$} & \colhead{log $N_{\rm H}$} & \colhead{$\frac{n_{\mathrm{C\,V}}}{n_{\mathrm{C\,IV}}}$\tablenotemark{b}} & \colhead{$t_r$\tablenotemark{c}} & \colhead{$R$} \\
\colhead{} & \colhead{($km\,s^{-1}$)} & \colhead{($km\,s^{-1}$)} & \colhead{} & \colhead{(cm$^{-2}$)} & \colhead{} & \colhead{(days)} & \colhead{(pc)}
}
\startdata
1 & 1162.9 & $1394.2 - 856.0$ & $-1.81^{+0.13}_{-0.12}$ & $19.40^{+0.06}_{-0.06}$ & $0.579^{+0.125}_{-0.046}$ & $< 2.51$ & $< 3.80$ \\
2 & 789.0 & $856.0-750.0$ & $-2.63^{+0.30}_{-0.34}$ & $17.74^{+0.42}_{-0.27}$ & $0.101^{+0.000}_{-0.000}$ & $36.06 \pm 2.23$ & $43.83^{+15.73}_{-12.27}$ \\
3 & 652.1 & $750.0-547.0$ & $-2.56^{+0.10}_{-0.11}$ & $18.45^{+0.10}_{-0.08}$ & $0.099^{+0.003}_{-0.001}$ & $37.15 \pm 1.50$ & $36.63^{+4.03}_{-3.88}$ \\
4 & 458.2 & $547.0-358.0$ & $-1.53^{+0.02}_{-0.02}$ & $18.49^{+0.02}_{-0.02}$ & $1.396^{+0.049}_{-0.022}$ & $< 2.51$ & $< 2.29$ \\
5 & 288.2 & $358.0- 120.0$ & $-1.00^{+0.03}_{-0.04}$ & $19.74^{+0.06}_{-0.06}$ & $4.934^{+0.943}_{-0.133}$ & $4.41 \pm 0.67$ & $0.97^{+0.26}_{-0.19}$ \\
6 & 44.3 & $120.0-0.0$ & $-0.56^{\infty}_{-0.71}$ & $19.44^{+1.34}_{-\infty}$ & $15.977^{+1.485}_{-0.270}$ & $4.83\pm 1.28$ & $2.81^{+2.92}_{-2.81}$ \\
\enddata
\tablecomments{Physical properties of the six UV absorption components in NGC 5548 derived using the D22 SED. }
\tablenotetext{a}{Velocity centroid of the component.}
\tablenotetext{b}{The ratio of the number densities of \ion{C}{5} to \ion{C}{4}.}
\tablenotetext{c}{The recombination timescale.}
\end{deluxetable*}

\section{Discussion}

\subsection{Origin of Broad Absorption Line Variability}
\label{sec:5.1}
The variability of absorption lines in AGNs is commonly attributed to two primary mechanisms: (1) changes in the ionizing continuum incident on the absorbing gas, and (2) transverse motion of the gas into or out of the line of sight, or a combination of both. In this work, we infer the distances of outflow components from their variability timescales, under the assumption that incident ionizing continuum changes is the dominating mechanism. Here, we assess the potential impact of the alternative mechanism (i.e. transverse motion of the absorber).

NGC 5548 is one of the most extensively studied Seyfert galaxies. The spectroscopic dataset used in this work is obtained during a reverberation mapping campaign taking it as a target (\cite{AGN_STORM}). According to the AGN STORM follow-up analysis (\cite{AGN_STORM_2021}), the C IV BLR extends over 20 light-days ($5.18 \times 10^{16}$ cm), and the central SMBH has a mass of $M_\mathrm{BH} = 7 \times 10^7\,M_\odot$. 
The smallest inferred outflow radius in our sample (component 5) is 0.97 pc. Assuming Keplerian motion, the transverse velocity at this radius is $v_\perp = 5.59 \times 10^7\ \mathrm{cm\,s^{-1}}$, implying that approximately $1.86 \times 10^9$ sec ($\sim 58.9$ years) be taken for the gas to traverse the BLR. This timescale is orders of magnitude longer than any measured recombination timescale, rendering transverse motion irrelevant on our observing timescales. Therefore, at the outflow radius determined above, the effect of transverse motion of the absorbing gas can be neglected.

Next, without assuming a priori distances, we attempt to assess the feasibility that the observed variability is caused by transverse motion of the gas. First, we note that the detection rate curves we obtained generally reach relatively high values (except for Trough I). Since the vertical axis of the detection rate curve represents detection probability and each bin contains dozens of samples, this indicates that variability is nearly ubiquitous. In contrast, a single event of an individual cloud moving into or out of the line of sight is unlikely to account for the frequent variability observed across multiple intervals. \textbf{Similar cases observed in typical Seyfert 1 galaxies—where UV absorption variability has been attributed to transiting clouds—indicate that such variability is generally intermittent or infrequent (e.g., NGC 4151, NGC 3516; \cite{2006ApJS..167..161K,2018ApJ...854..166D}).} Therefore, we tentatively interpret the observed recombination timescales as corresponding to Keplerian orbital periods. Under this assumption, $r = \left(\frac{T}{2\pi}\right)^{2/3} (GM)^{1/3}$, and the derived maximum radius among these outflow components is 0.000436 pc. However, reverberation mapping of NGC 5548 yields a BLR radius of about 20 light-days (0.0168 pc). In this case, the inferred outflow components would be located well within the BLR, which is inconsistent with observations.
If, alternatively, the observed variability timescales were attributed to a sequence of clouds—or inhomogeneous cloudlets—moving in and out of the line of sight, they would need to form a continuous, ring-like, highly inhomogeneous absorbing structure. Since the transverse velocity of the absorbing gas cannot be reliably constrained, we adopt the Keplerian velocity at the BLR radius ($4.25\times 10^{8}\ \mathrm{cm,s^{-1}}$) as an estimate. Multiplying this velocity by the measured recombination timescales yields rough estimates of the inhomogeneity scales of these outflow components. The maximum value obtained is $4.42\times 10^{-4}$ pc. Given that the C IV absorption outflows are located beyond the BLR, where the corresponding Keplerian velocity is lower, the degree of inhomogeneity would be even larger. At present, however, there is no observational evidence supporting the existence of such a continuous, ring-like absorbing structure.

Other sporadic mechanisms that could in principle drive absorption line variability may also be considered.
For example, instabilities within the outflows themselves (with timescales comparable to the dynamical scales discussed above) or the natural dissipation of clouds without shearing or external forces (with a sound-crossing timescale of $\sim$8000 years for a cloud of radius 0.1 pc and temperature $10^4$ K) require timescales far longer than those measured in this work and are therefore ruled out (\cite{2013MNRAS.429.1872C}). Moreover, magnetically driven outflows are expected to accelerate the gas only to a few times the Keplerian velocity (\cite{2014ApJ...780..120F}), which, as discussed above, does not affect our conclusions.

Previous works have also identified an ``obscuring wind'' in NGC 5548, which may block up to 90\% of the soft X-ray flux and enabling low-ionization species to persist, leading to broad UV absorption troughs (\cite{Kaastra_2014_ngc5548_2013_campaign}). Such a wind could potentially introduce variability through changes in covering fraction, especially in the C IV troughs. However, its effects are observed on decade-long timescales (\cite{Mehdipour_2024}), whereas our dataset spans about two months in early 2014, with supplementary post-June 2013 spectra. Therefore, the obscuring wind is unlikely to affect the variability measured here.

\begin{figure}
    \centering
    \includegraphics[width=0.75\linewidth]{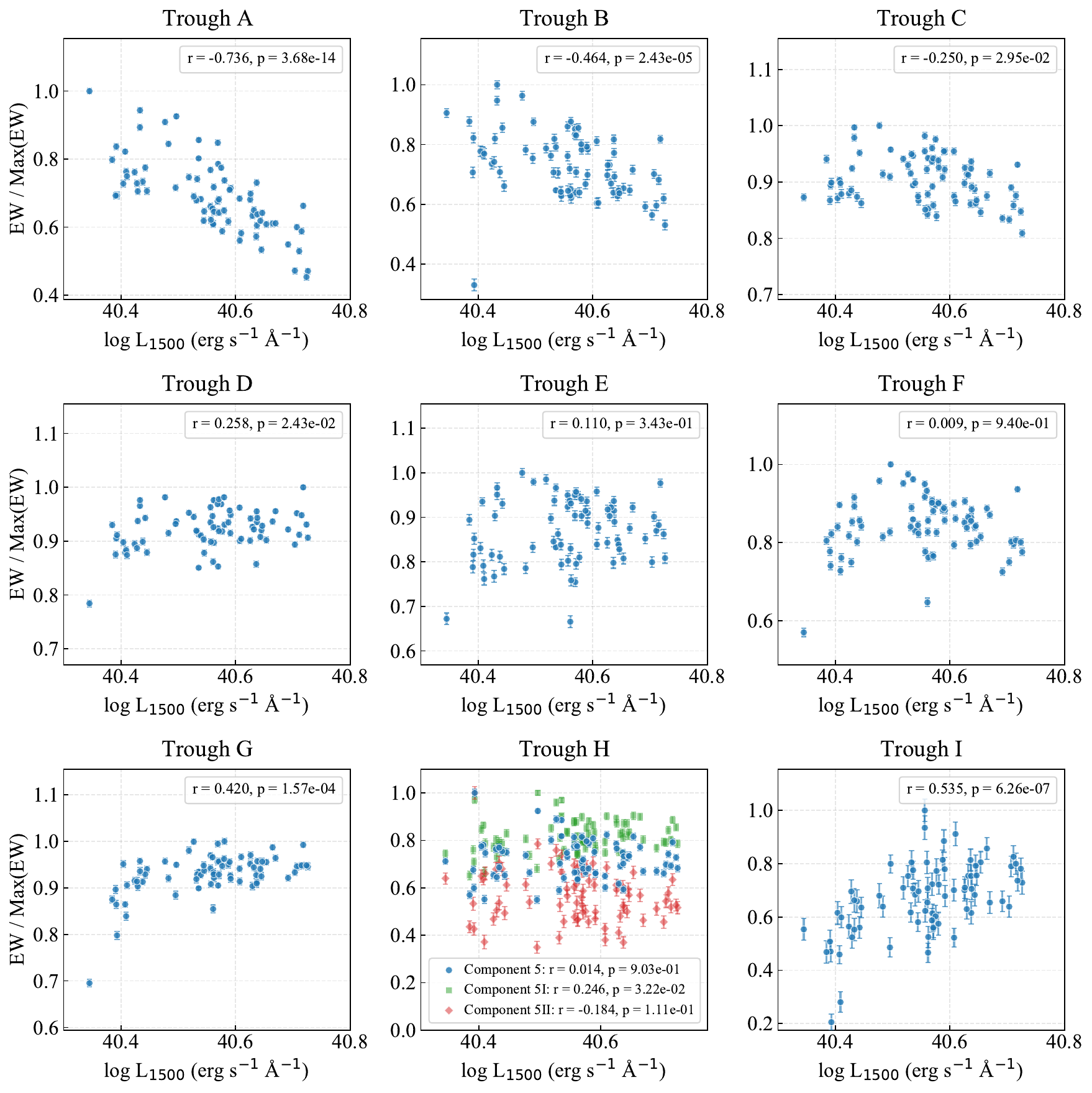}
    \caption{\textbf{Correlation between the 1500 Å continuum luminosity and the equivalent width (EW) of the C IV absorption troughs in NGC 5548.} 
    Each panel corresponds to a distinct absorption trough, with the Spearman rank correlation coefficient r and associated p-value indicated. The red and green points in trough H represent the decomposed components 5I and 5II, obtained by splitting the original velocity component 5. Their respective positive and negative trends help explain the near-zero overall correlation initially observed for component 5.
    A significant anti-correlation is observed only in Trough A (component 1), where the recombination timescale is substantially shorter than the observational cadence. In other troughs, the absence of a clear correlation does not necessarily imply a different variability mechanism. Instead, when the sampling interval approaches or falls below the recombination timescale, the absorbing gas may not have sufficient time to respond to continuum fluctuations, thereby weakening the observed correlation.}
    \label{fig:L1500-EW}
\end{figure}

To further verify our assumption, we examine the correlation between the 1500 Å continuum luminosity and the EW of the C IV absorption in each trough using Spearman’s rank correlation (Figure~\ref{fig:L1500-EW}, \cite{He_2017,He_2022}). When the observing interval is significantly longer than the recombination timescale of gas, this correlation analysis provides a straightforward test of continuum-driven variability. However, it becomes less reliable when the interval approaches or falls below $t_r$. As expected, only trough A (component 1) shows a significant anti-correlation (Spearman coefficient $r = -0.736$). Other troughs exhibit weak correlations, the strongest being 0.535. Notably, trough H (component 5) shows virtually no correlation ($r = 0.014$), whereas trough I (component 6), which also has a $t_r$ on the order of days, displays the strongest correlation after trough A ($r = 0.535$).

Inspection of the spectra suggests that trough H may include a secondary sub-component on its red side. \cite{Crenshaw_2003} reported a velocity gap between components 5 and 6. Based on this, we divide component 5 at $-261.0$ km s$^{-1}$ into two sub-components, designated 5I and 5II, and overplot their respective 1500 Å continuum flux and EW correlations in the trough H panel. Notably, these sub-components exhibit opposite correlation signs: one positive, one negative. 
This suggests that the original component 5 may in fact be a blend of two outflow components with different ionization states, leading to a negligible overall correlation. 
However, photoionization modeling with \texttt{CLOUDY} yields ionization parameters of $U = -1.08^{+0.04}_{-0.05}$ for 5I and $U = -0.97^{+0.11}_{-0.08}$ for 5II, hydrogen column densities of $\log N_H = 19.40^{+0.09}_{-0.11}$ and $19.30^{+0.15}_{-0.10}$, and ion ratios $n_{\mathrm{C\,V}}/n_{\mathrm{C\,IV}}$ of $3.810^{+0.502}_{-0.083}$ and $5.086^{+0.521}_{-0.099}$, respectively, manifesting a difference less significant than expected.

\subsection{Results Based on the Obscured SED}
In this work, we adopt the D22 SED (\cite{D22}) as our default ionizing spectral energy distribution. This SED represents the intrinsic, unabsorbed continuum of NGC 5548, reconstructed from the same 2014 spectroscopic dataset used in our analysis. Our choice is motivated by two key considerations: 1) it is directly based on the same dataset employed in this work, whereas the obscured SED derived from that campaign suffers from a gap over $10^{15.2}$--$10^{17.0}$ Hz; 2) previous works have demonstrated that the obscuring wind in NGC 5548 is spatially inhomogeneous, introducing non-uniform attenuation across different outflow components (\citealt{Mehdipour_2024}), making it uncertain whether all components are equally affected.

Nonetheless, for the purpose of comparison, we also perform photoionization modeling using the obscured SED reconstructed by \citet{Mehdipour_2015_ngc5548_2013_campaign} based on multiwavelength observations conducted in 2013.
Adopting this obscured SED, we estimate a bolometric luminosity of $L_\mathrm{bol} = 3.55 \times 10^{47},\mathrm{erg\,s^{-1}}$ based on the average continuum flux at 1500 Å measured from the composite spectrum. 
This value is significantly higher than those reported in the literature, e.g., $L_\mathrm{bol} = 2.82 \times 10^{44},\mathrm{erg\,s^{-1}}$ in \citet{Ebrero_2016}, $2.6 \times 10^{44},\mathrm{erg\,s^{-1}}$ in \citet{Arav_2015}, and $1.7 \times 10^{44},\mathrm{erg\,s^{-1}}$ in \citet{2007ApJ...662..205B}. Consequently, the derived ionizing photon rate, $Q_H = 1.48 \times 10^{55},\mathrm{s^{-1}}$, is also about an order of magnitude higher than that in the D22 SED.

The resultant photoionization parameters for each velocity component are given below:
\begin{itemize}
    \item $\log U_{\mathrm{H}}$: $-2.24^{+0.09}_{-0.08}, -2.77^{+0.17}_{-0.17}, -2.73^{+0.05}_{-0.06}, -2.04^{+0.02}_{-0.02}, -1.64^{+0.03}_{-0.03}, -1.25^{+\infty}_{-0.60}$
    \item $\log N_\mathrm{H}$: $19.46^{+0.05}_{-0.06}, 17.61^{+0.17}_{-0.15}, 18.36^{+0.04}_{-0.03}, 18.53^{+0.01}_{-0.01}, 19.65^{+0.05}_{-0.05}, 19.24^{+1.70}_{-\infty}$
    \item $n_{\mathrm{C\,V}}/n_{\mathrm{C\,IV}}$: $0.753^{+0.189}_{-0.052}, 0.185^{+0.000}_{-0.000}, 0.233^{+0.005}_{-0.002}, 1.486^{+0.053}_{-0.026}, 4.204^{+0.729}_{-0.120}, 14.580^{+0.845}_{-0.205}$
    \item Derived radial distances $R$ (pc): $< 18.73$, $146.19^{+25.42}_{-24.16}$, $133.82^{+7.89}_{-8.11}$, $< 12.64$, $3.06^{+2.78}_{-1.02}$, $17.56^{+14.64}_{-17.56}$.
\end{itemize}

The full parameter set is provided in Table~\ref{table:obscured_SED}. We conclude that the radial distance of outflows derived under the obscured SED is systematically larger than that obtained with the D22 SED (typically by about an order of magnitude).

\subsection{Comparison with Previous Studies}
\begin{figure}
    \centering
    \includegraphics[width=0.75\linewidth]{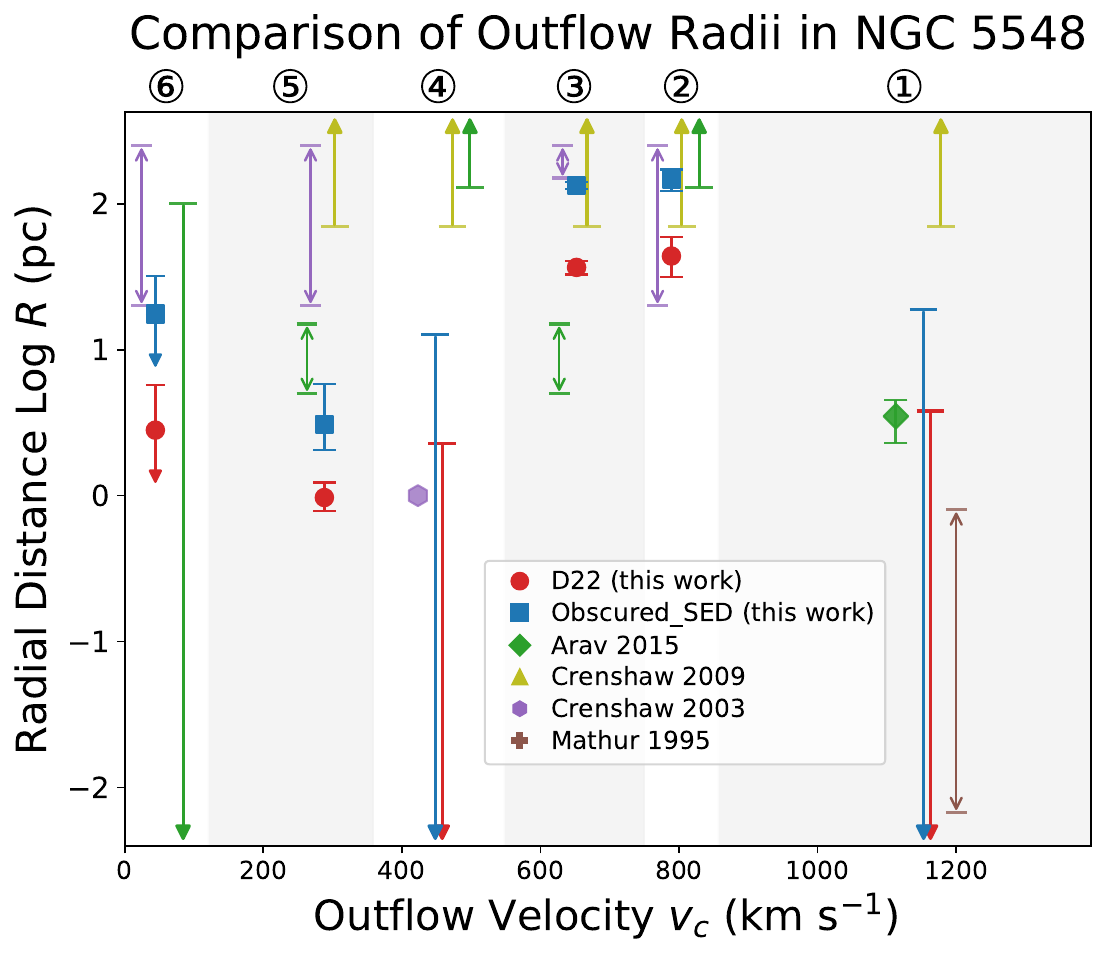}
    \caption{\textbf{Comparison of UV outflow radial distances in NGC 5548 measured in this work and previous studies.} The horizontal axis shows the central velocity $v_c$ of each outflow component. The vertical axis shows the logarithmic radial distance $R$ (in parsecs). The numeric labels above each shaded band indicate the corresponding velocity component, and the alternating shaded regions mark the velocity ranges of each component. Results from this work are shown in red circles (D22 SED) and blue squares (obscured SED). Data from the literature are plotted with different symbols and colors, as indicated in the legend. Arrows indicate upper or lower limits and range constraints, with their colors matching the corresponding study.}
    \label{fig:comparison}
\end{figure}

We compare the radial distance of the UV outflow components in NGC 5548 measured in this work with previous works, as summarized in a velocity-radius diagram (Figure~\ref{fig:comparison}).

The division of velocity components for the ultraviolet outflows adopted in this work, as well as in \citet{Arav_2015} and \citet{Crenshaw_2009}, follows the scheme established by \citet{Crenshaw_2003}. Due to differences in the adopted systemic redshift and the absorption trough profiles across different datasets, the central velocities and velocity ranges of these components vary slightly across the literature. Our visual inspection of the involved spectra suggest that the comparison of the outflow components being compared here are based on the same structures. In Figure~\ref{fig:comparison}, outflow components are therefore plotted using the velocity centroids defined in this work, but slight horizontal offsets are used to avoid overlapping symbols.

In most previous works, the lack of detectable excited-state absorption lines, which are instrumental to electron density measurement, posed a limitation to the available methods to derive outflow distances. As a result, distances are typically estimated by examining whether absorption lines respond to continuum variations over different epochs, providing rough constraints on recombination timescales and, consequently, the outflow locations. An exception is \citet{Arav_2015}, where the authors identify a remarkable amount of previously undetected ionic absorption troughs associated with component 1 in the 2013 HST/COS spectrum of NGC 5548, allowing them to use excited-state ionic column density ratios to derive a reliable distance for component 1.

In \citet{Mathur_1995}, the authors propose that the UV and X-ray absorbers in NGC 5548 are manifestations of the same physical outflow. By analyzing the absorption characteristics of Ly$\alpha$, N V, C IV, O VII, O VIII, and Fe XVII in both X-ray and UV spectra, they estimate the physical properties of the absorber and infer an upper limit on the radial distance of the C IV absorber based on its response time. Furthermore, the observed C IV absorption depth exceeding the continuum minimum during their observations imply that the absorber at least partially covers the C IV emission region, providing a lower limit on its distance. 
However, subsequent investigations have shown that while UV outflows and warm absorbers (WAs) exhibit kinematic associations, they may represent physically distinct structures, and WAs show higher ionization parameters in general (\cite{2002ApJ...565..800B, Crenshaw_2003, Arav_2015}).
Therefore, the radial distances of the UV outflows reported by \citet{Mathur_1995} are likely underestimated.

\citet{Crenshaw_2003} also discuss earlier works in which narrow emission-line profile fits suggested that component 4 does not cover the narrow line region (NLR), as the fitted narrow line flux barely reach the bottom of the component 4 absorption trough. However, by analyzing low-state spectra, in which the UV continuum and broad emission-line fluxes are at low levels, \citet{Crenshaw_2003} find that in all three ionic absorption species (Ly$\alpha$, N V, and C IV), the narrow-line flux exceed the bottom of the component 4 absorption trough. This conclusion indicate that component 4 at least partially covers the inner high-ionization narrow-line region located $\sim$1 pc from the central ionizing source.
Assuming that component 4 has been transiting the NLR during that time, the absorber would have moved roughly 0.0056 pc over the 12 years by 2014 (the year the data  analyzed in this work was taken). This distance is consistent with what we infer for this component ($<2.29$ pc).

\citet{Crenshaw_2009} report that during a period of UV continuum dimming from 2002 to 2004, the absorbers show no response, providing an upper limit on the recombination timescale of 2.0 years. Based on this, they conclude that the UV outflows are located at distances greater than 70~pc.

The comparison of our UV outflow distance measurements (under both adopted SEDs) to the literature is visualized in Figure~\ref{fig:comparison}. Data points from the same work are shown in identical symbols and colors. If the results are expressed as a range, arrows of the same color indicate the corresponding boundaries.
It can be seen that the majority of UV outflow distance estimates for NGC 5548 in the literature are given as broad ranges and show considerable scatter. Nevertheless, our results agree well with the more reliable measurements in the literature, namely, component 1 from \citet{Arav_2015} and component 4 from \citet{Crenshaw_2003}. Component 2 and 3 are consistently placed at relatively large distances by all works. As for component 6, \citet{Arav_2015} report a broad upper limit ($R < 100$~pc), within which our measurement also falls.

We further emphasize that, except for component 1 in \citet{Arav_2015} and component 4 in \citet{Crenshaw_2003}, all other distance estimates are based on constraints from the recombination timescale of the absorbing gas. The dataset used in this work consists of high-resolution HST/COS spectra from the AGN STORM campaign, obtained with an unprecedented cadence of nearly one spectrum per day over a six-month monitoring period (although only about two and a half months are used in our analysis). Compared to previous works that typically obtain only a few spectra over several years, this dataset provides a significant advantage in detecting absorption-line variability and constraining recombination timescales, owing to its high sampling rate and continuous time coverage.

Nevertheless, even with this advantageous dataset, the detection rate curves for the C IV absorption troughs alone are insufficient to determine precise distances for all UV outflow components. This is primarily because the recombination timescales of component 1 and 4 remain shorter than the typical observational intervals in the AGN STORM dataset, and component 4 is also affected by blending from other velocity components. 

\subsection{Impact of Column Density and Ionization Parameter on the Results}

The ionization parameters adopted in this work are derived from column densities obtained through partial covering fits to the absorption troughs, combined with Cloudy photoionization simulations (see Appendix~\ref{appendix:cloudy}). Inevitably, effects such as hidden saturation and blending of multiple velocity components in the troughs can influence the inferred column densities, and constraining the models using only the available Si IV and C IV lines may introduce additional biases. Unfortunately, our spectra contain only these two usable absorption lines, limiting us to a relatively coarse estimate.

On the positive side, unlike the traditional method based on ionic excited states, in our approach the measurement of column densities affects only the value of the ionization parameter $U_\mathrm{H}$. According to Equation~\ref{eq:2}, the impact of $U_\mathrm{H}$ scales approximately as $\sim U_\mathrm{H}^{-0.5}$. Thus, even if $U_\mathrm{H}$ were misestimated by an order of magnitude, the resulting variation in outflow distance $R$ would be within a factor of $0.3$–$3$, corresponding to $\Delta \log R = \pm 0.5$ in logarithmic space.

To roughly estimate the potential deviation of our measured ionization parameters, we refer to the results of \cite{Arav_2015}. According to their analysis, the ionization parameter of the NGC 5548 outflow decreased gradually from year to year (see their Fig. 5). Importantly, \cite{Arav_2015} analyzed the 2013 spectra and provided the only measurement of the outflow distance in NGC 5548 using the traditional ionic excited-state method. Based on P III, P V, Fe III, and Si II absorption lines, they derived $\log(U_\mathrm{H}) = -1.5^{+0.4}_{-0.2}$ for component 1. For component 3, they obtained $\log(U_\mathrm{H}) = -2 \sim -1.1$ using C II, Si III, Si IV, and N V, while for component 5, they derived $\log(U_\mathrm{H}) = -1.8 \sim -1.2$. Other components could not be constrained due to the lack of available absorption lines.
In comparison, our measurements yield $\log(U_\mathrm{H}) = -1.81^{+0.13}_{-0.12}$ for component 1, $\log(U_\mathrm{H}) = -2.56^{+0.10}_{-0.11}$ for component 3, and $\log(U_\mathrm{H}) = -1.00^{+0.03}_{-0.04}$ for component 5. Among these, only component 3 shows a significant discrepancy ($\Delta \log U_\mathrm{H} \sim -0.5$). In this case, the corresponding difference in outflow radius is $\Delta \log R \sim 0.25$, which remains modest.

\textbf{Furthermore, considering the possible effects of line saturation, we treat the ionic column densities of C IV and Si IV derived from the spectra as lower limits. Using these values, we obtained a corresponding set of lower limits on the ionization parameter $U_H$ from our CLOUDY simulations. Consequently, we can also infer approximate upper limits on the outflow distances $R$, as summarized in Table~\ref{tab:saturation}. As shown in the table, under both adopted SEDs, the upper limits of $R$ for most velocity components are only about two to four times larger than the values derived from the nominal column densities. Except for component 6, all upper limits of $R$ remain below 6.3 times their corresponding precise measurements. Therefore, our previous conclusions remain unaffected.}

\section{Conclusion}
In this work, we measure the radial distances of UV outflow components in NGC 5548 by analyzing absorption-line variability using high-cadence HST/COS UV spectra from the AGN STORM campaign, supplemented by archival data taken in 2013. We employ the detection rate curve method to characterize the recombination timescale ($t_r$) for each outflow component, and combine the outcome with photoionization modeling from \texttt{CLOUDY} to derive their ionization parameters ($U_H$), hydrogen column densities ($N_H$), ion density ratio ($n_{C\,V}/n_{C\,IV}$), so that radial distances ($R$) are derived or constrained. The main conclusions of this work are as follows:
\begin{itemize}
    \item The detection rate curve method are used to infer the recombination timescales of several velocity components with success. For components 2, 3, 5, and 6, detection rate profiles allow robust $t_r$ measurements, ranging from $\sim4$ to $\sim40$ days. For components 1 and 4, the absence of detectable trends in the detection rate curves implies recombination timescales shorter than the typical observational interval of 2.51 days, placing upper limits on their radial distances.
    \item The derived radial distances of the UV outflow components span from sub-parsec to tens-of-parsec scales. The innermost components, 1, 4, 5, and 6, are located within $\sim$1--3 pc, while components 2 and 3 reside at distances beyond 30 pc.
    \item Two absorption troughs (E and F) exhibit distinct ``double-step'' features in their detection rate curves, allowing us to disentangle the blended contributions from multiple velocity components. In particular, the longer $t_r$ components inferred from these curves correspond to components 2 and 3, which are also associated with the largest radial distances in our sample, reaching $\sim44$ pc and $\sim37$ pc, respectively.
    \item Compared to previous works, our distance estimates for the UV outflow components in NGC 5548 show overall consistency, despite the considerable scatter in the literature. Our results agree well with the more reliable results in the literature on components 1 and 4. Owing to the high-cadence, regularly sampled HST/COS spectra, our method provides improved sensitivity to absorption-line variability and yields more reliable constraints on recombination timescales and outflow distances.
\end{itemize}

\begin{deluxetable*}{cccccccc}
\tablecaption{Physical Parameters of the C\,\textsc{iv} Components in the Obscured SED \label{table:obscured_SED}}
\tablehead{
\colhead{Component} & \colhead{$v_c$\rlap{\tablenotemark{a}}} & \colhead{Velocity Range} & \colhead{log $U_{\mathrm{H}}$} & \colhead{log $N_{\rm H}$} & \colhead{$\frac{n_{\mathrm{C\,V}}}{n_{\mathrm{C\,IV}}}$\tablenotemark{b}} & \colhead{$t_r$\tablenotemark{c}} & \colhead{$R$} \\
\colhead{} & \colhead{($km\,s^{-1}$)} & \colhead{($km\,s^{-1}$)} & \colhead{} & \colhead{(cm$^{-2}$)} & \colhead{} & \colhead{(days)} & \colhead{(pc)}
}
\startdata
1 & 1162.9 & $1394.2-856.0$ & $-2.24^{+0.09}_{-0.08}$ & $19.46^{+0.05}_{-0.06}$ & $0.753^{+0.189}_{-0.052}$ & $< 2.51$ & $< 18.73$ \\
2 & 789.0 & $856.0-750.0$ & $-2.77^{+0.17}_{-0.17}$ & $17.61^{+0.17}_{-0.15}$ & $0.185^{+0.000}_{-0.000}$ & $36.06 \pm 2.23$ & $146.19^{+25.42}_{-24.16}$ \\
3 & 652.1 & $750.0-547.0$ & $-2.73^{+0.05}_{-0.06}$ & $18.36^{+0.04}_{-0.03}$ & $0.233^{+0.005}_{-0.002}$ & $37.15 \pm 1.50$ & $133.82^{+7.89}_{-8.11}$ \\
4 & 458.2 & $547.0-358.0$ & $-2.04^{+0.02}_{-0.02}$ & $18.53^{+0.01}_{-0.01}$ & $1.486^{+0.053}_{-0.026}$ & $< 2.51$ & $< 12.64$ \\
5 & 288.2 & $358.0-120.0$ & $-1.64^{+0.03}_{-0.03}$ & $19.65^{+0.05}_{-0.05}$ & $4.204^{+0.729}_{-0.120}$ & $4.41 \pm 0.67$ & $3.06^{+2.78}_{-1.02}$ \\
6 & 44.3 & $120.0-0.0$ & $-1.25^{+inf}_{-0.60}$ & $19.24^{+1.70}_{-\infty}$ & $14.580^{+0.845}_{-0.205}$ & $4.83 \pm 1.28$ & $17.56^{+14.64}_{-17.56}$ \\
\enddata
\tablecomments{Physical properties of the six UV absorption components in NGC 5548 derived using the Obscured SED. 
}
\tablenotetext{a}{Velocity centroid of the component.}
\tablenotetext{b}{The ratio of the number densities of \ion{C}{5} to \ion{C}{4}.}
\tablenotetext{c}{The recombination timescale.}
\end{deluxetable*}

\begin{deluxetable*}{cccccc}
\tablecaption{Derived $U_H$ and $R$ from nominal and lower-limit $N_{\rm ion}$ for both SEDs.\label{tab:saturation}}
\tabletypesize{\small}     
\tablewidth{0pt}

\tablehead{
\colhead{Component} &
\colhead{Treatment of $N_{ion}$ \tablenotemark{a}} &
\multicolumn{2}{c}{\textbf{D22}} &
\multicolumn{2}{c}{\textbf{Obscured SED}} \\
\colhead{} & \colhead{} &
\colhead{log $U_{\mathrm{H}}$} & \colhead{$R$ (pc)} &
\colhead{log $U_{\mathrm{H}}$} & \colhead{$R$ (pc)}
}

\startdata
1 &  nominal      & $-1.81^{+0.13}_{-0.12}$ & $< 3.80$ & $-2.24^{+0.09}_{-0.08}$ & $< 18.73$ \\
  &  lower limits & $\geq -2.6$ & $\leq 8.86$ & $\geq -2.8$ & $\leq 34.00$ \\
\tableline
2 &  nominal      & $-2.63^{+0.30}_{-0.34}$ & $43.83^{+15.73}_{-12.27}$ & $-2.77^{+0.17}_{-0.17}$ & $146.19^{+25.42}_{-24.16}$ \\
  &  lower limits & $\geq -3.4$ & $\leq 90.13$ & $\geq -3.6$ & $\leq 351.21$ \\
\tableline
3 & nominal       & $-2.56^{+0.10}_{-0.11}$ & $36.63^{+4.03}_{-3.88}$ & $-2.73^{+0.05}_{-0.06}$ & $133.82^{+7.89}_{-8.11}$ \\
  & lower limits  & $\geq -3.2$ & $\leq 72.69$ & $\geq -3.3$ & $\leq 250.78$ \\
\tableline
4 & nominal       & $-1.53^{+0.02}_{-0.02}$ & $< 2.29$ & $-2.04^{+0.02}_{-0.02}$ & $< 12.64$ \\
  & lower limits  & $\geq -3.0$ & $\leq 12.23$ & $\geq -3.2$ & $\leq 47.34$ \\
\tableline
5 & nominal       & $-1.00^{+0.03}_{-0.04}$ & $0.97^{+0.26}_{-0.19}$ & $-1.64^{+0.03}_{-0.03}$ & $3.06^{+2.78}_{-1.02}$ \\
  & lower limits  & $\geq -2.6$ & $\leq 6.29$ & $\geq -2.8$ & $\leq 12.38$ \\
\tableline
6 & nominal       & $-0.56^{\infty}_{-0.71}$ & $2.81^{+2.92}_{-2.81}$ & $-1.25^{+inf}_{-0.60}$ & $17.56^{+14.64}_{-17.56}$ \\
  & lower limits  & $\geq -3.0$ & $\leq 36.71$ & $\geq -3.2$ & $\leq 135.98$ \\
\enddata

\tablenotetext{a}{Treatment of spectroscopic $N_{\mathrm{ion}}$ when comparing with CLOUDY model: nominal measurements vs.\ lower limits (to account for saturation); the resulting $U_H$ and $R$ are listed.}
\end{deluxetable*}

\begin{figure}
    \centering
    \includegraphics[width=1\linewidth]{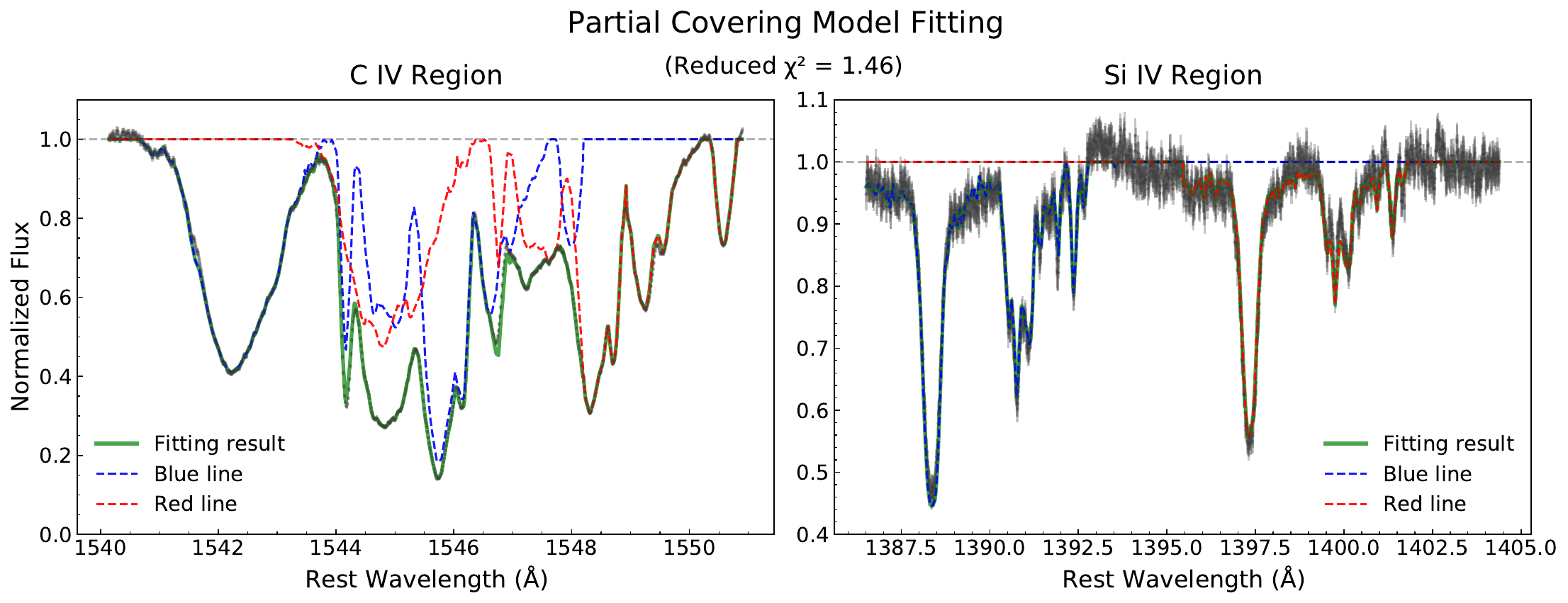}
    \caption{\textbf{Partial covering model fits to the normalized spectra of the \ion{C}{4} (left) and \ion{Si}{4} (right) absorption regions.} 
    The black curves show the relative fluxes of the C IV and Si IV absorption troughs with 1$\sigma$ uncertainties. 
    The best-fit model (solid green line) simultaneously fits both doublets, with individual contributions from the blue and red members indicated by dashed blue and red lines, respectively. 
    The reduced chi-squared value for the joint fit is $\chi^2_\nu = 1.46$.
    }
    \label{fig:partial-covering-model-fit}
\end{figure}

\begin{figure*}[htbp]
    \centering
    \includegraphics[width=0.32\textwidth]{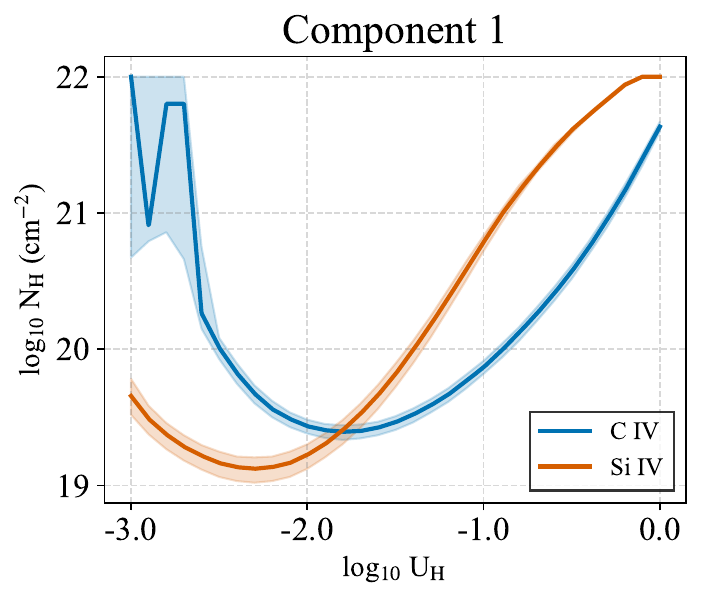}
    \includegraphics[width=0.32\textwidth]{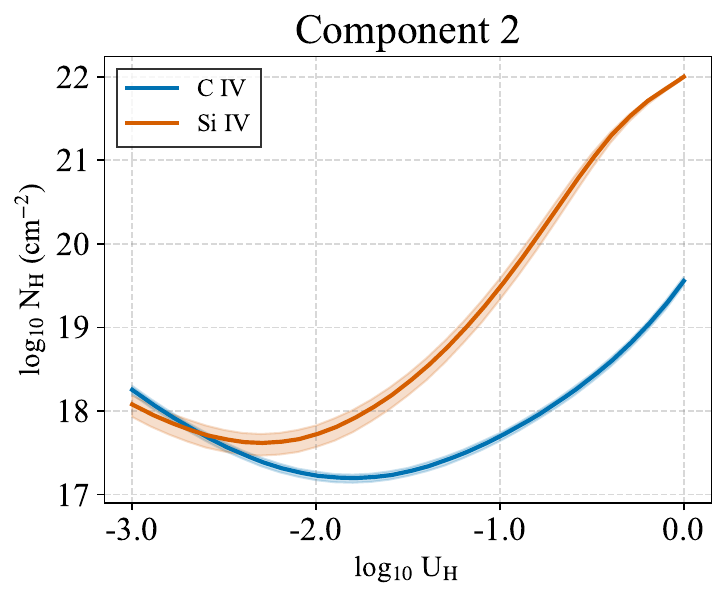}
    \includegraphics[width=0.32\textwidth]{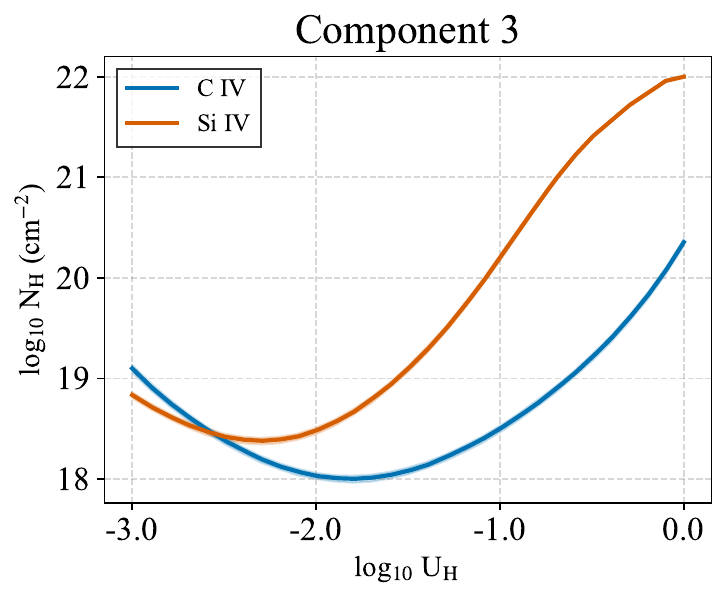}\\
    \includegraphics[width=0.32\textwidth]{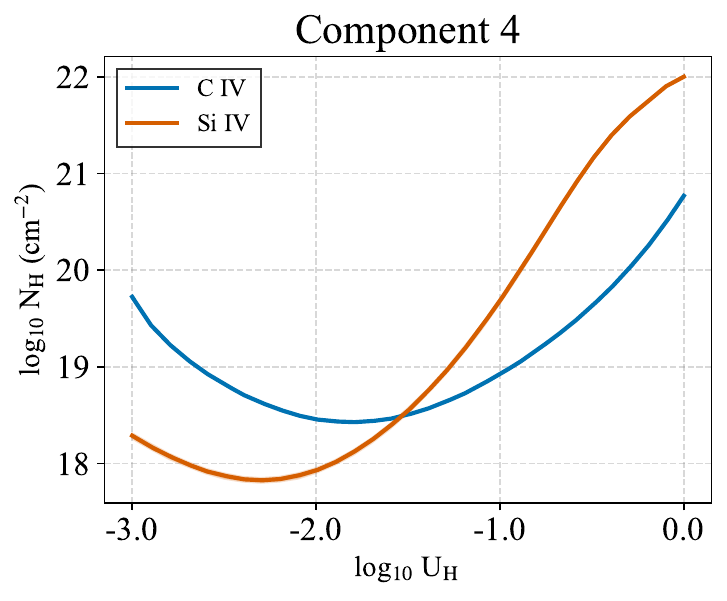}
    \includegraphics[width=0.32\textwidth]{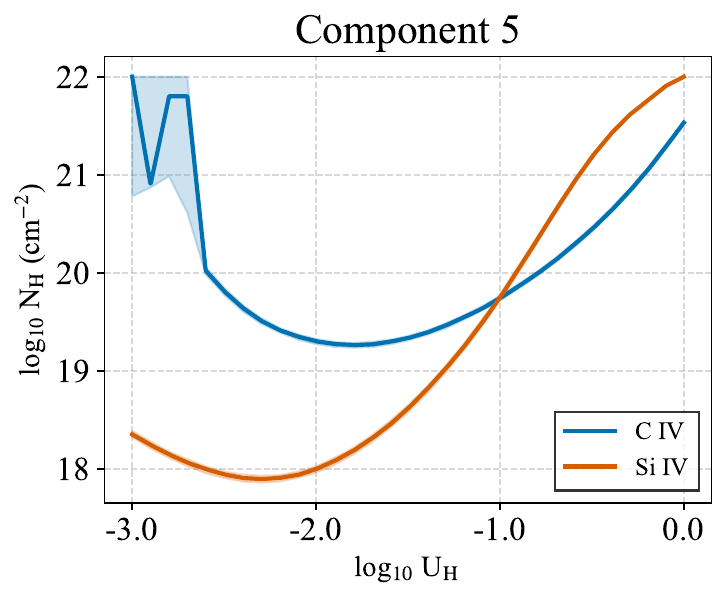}
    \includegraphics[width=0.32\textwidth]{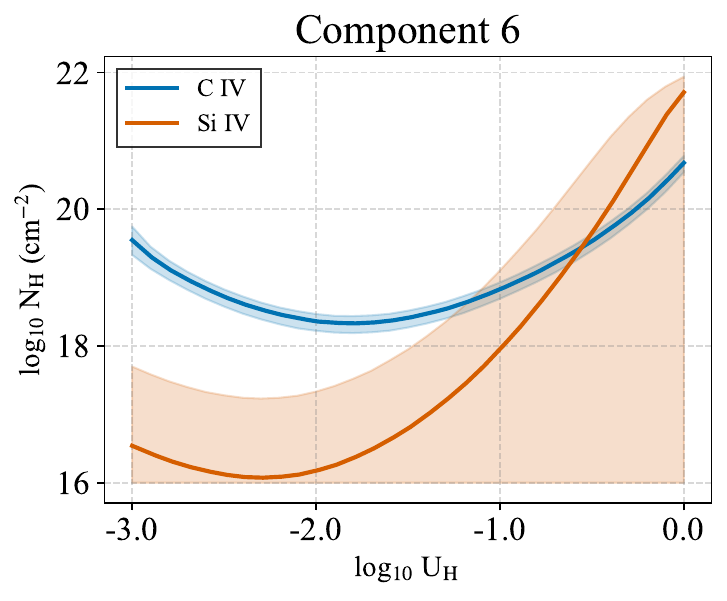}
    \caption{\textbf{CLOUDY-based photoionization modeling of outflow components using the D22 SED.}
    Constraints on the ionization parameter $U_{\mathrm{H}}$ and hydrogen column density $N_{\mathrm{H}}$ for the six velocity-defined outflow components in NGC 5548, based on photoionization modeling using the D22 SED. Each panel shows the $\log U_{\mathrm{H}}$–$\log N_{\mathrm{H}}$ parameter space computed from a grid of CLOUDY simulations for a given component, with model-predicted column densities compared to observations to identify the allowed region for each ion. The colored bands represent the 1$\sigma$ allowed regions for the observed ionic column densities of \ion{C}{4} (blue) and \ion{Si}{4} (orange), as derived from partial covering model fits. The intersection point of the two bands is adopted as the best-fit solution for each component.
    }
    \label{fig:cloudy_D22}
\end{figure*}

\begin{figure*}[htbp]
    \centering
    \includegraphics[width=0.32\textwidth]{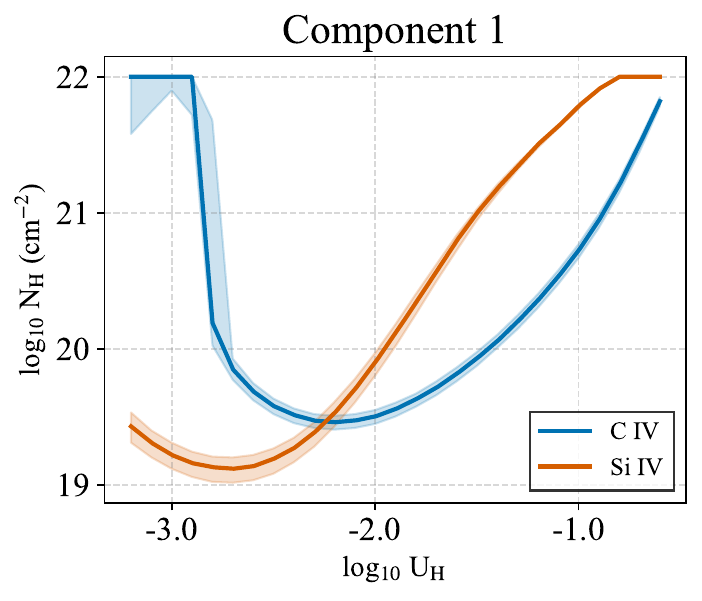}
    \includegraphics[width=0.32\textwidth]{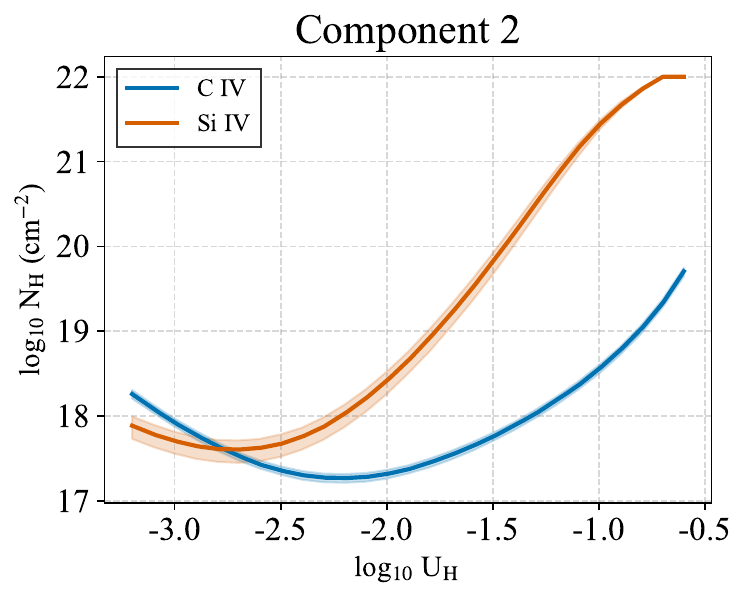}
    \includegraphics[width=0.32\textwidth]{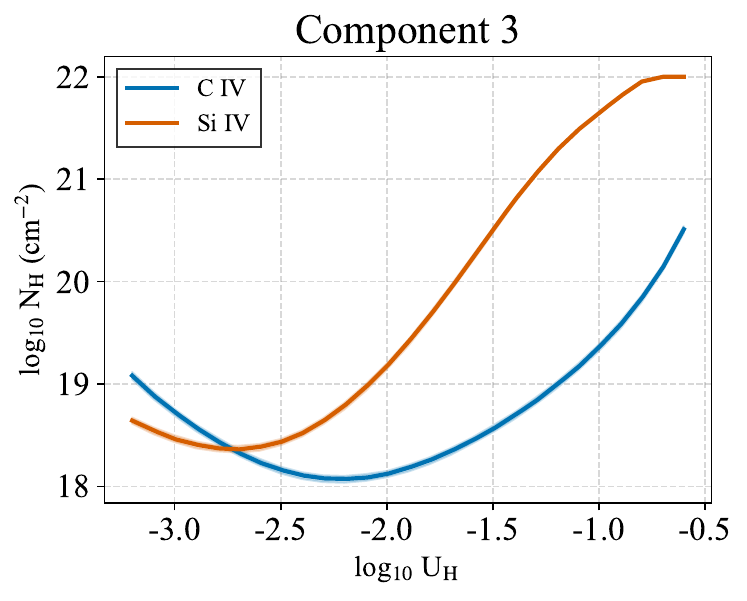}\\
    \includegraphics[width=0.32\textwidth]{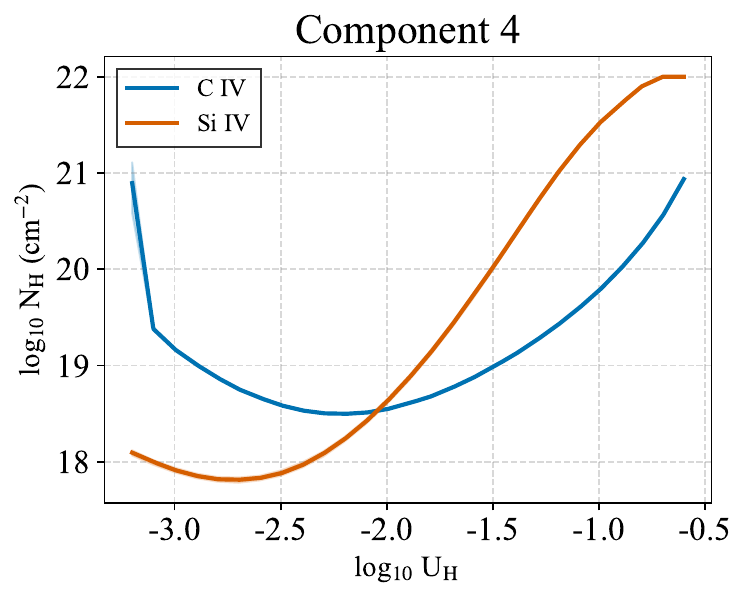}
    \includegraphics[width=0.32\textwidth]{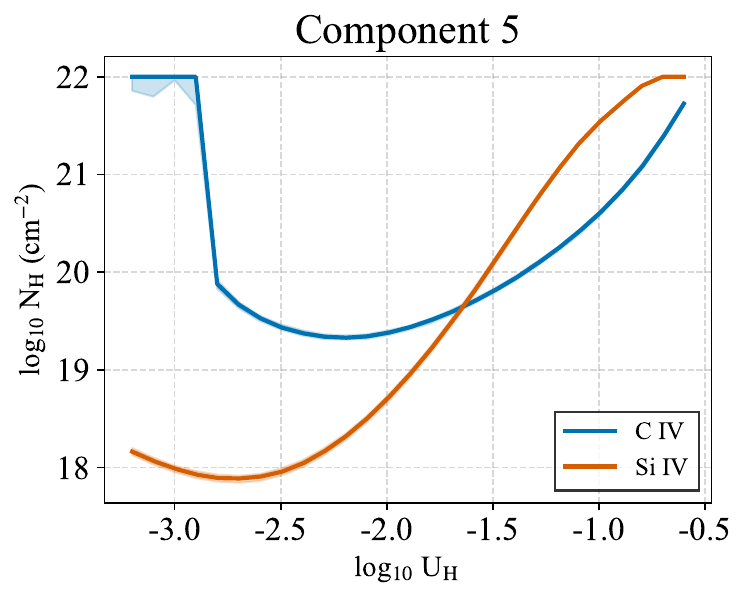}
    \includegraphics[width=0.32\textwidth]{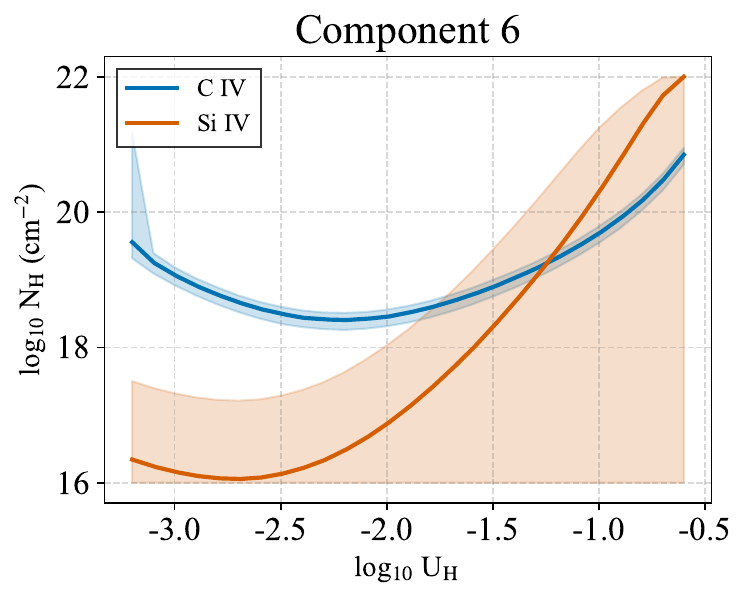}
    \caption{\textbf{CLOUDY-based photoionization modeling of outflow components using the NGC 5548 Obscured SED.}
    Constraints on the ionization parameter $U_{\mathrm{H}}$ and hydrogen column density $N_{\mathrm{H}}$ for the six velocity-defined outflow components in NGC 5548, based on photoionization modeling using the NGC 5548 Obscured SED. Each panel shows the $\log U_{\mathrm{H}}$–$\log N_{\mathrm{H}}$ parameter space computed from a grid of CLOUDY simulations for a given component, with model-predicted column densities compared to observations to identify the allowed region for each ion. The colored bands represent the 1$\sigma$ allowed regions for the observed ionic column densities of \ion{C}{4} (blue) and \ion{Si}{4} (orange), as derived from partial covering model fits. The intersection point of the two bands is adopted as the best-fit solution for each component.
    }
    \label{fig:cloudy_obscuredSED}
\end{figure*}

\begin{figure}[htbp]
    \centering
    \includegraphics[width=0.95\textwidth]{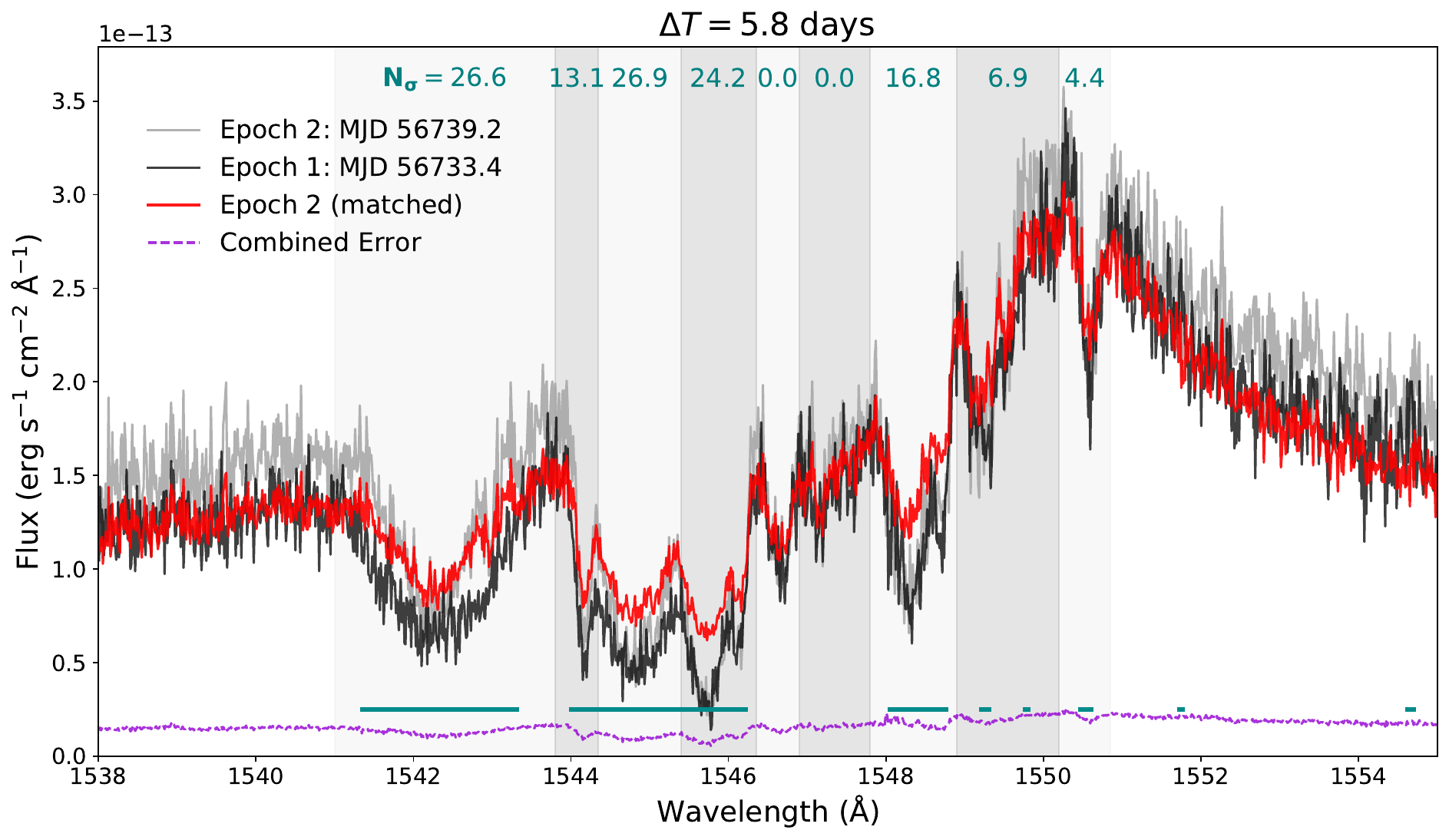}
    \includegraphics[width=0.95\textwidth]{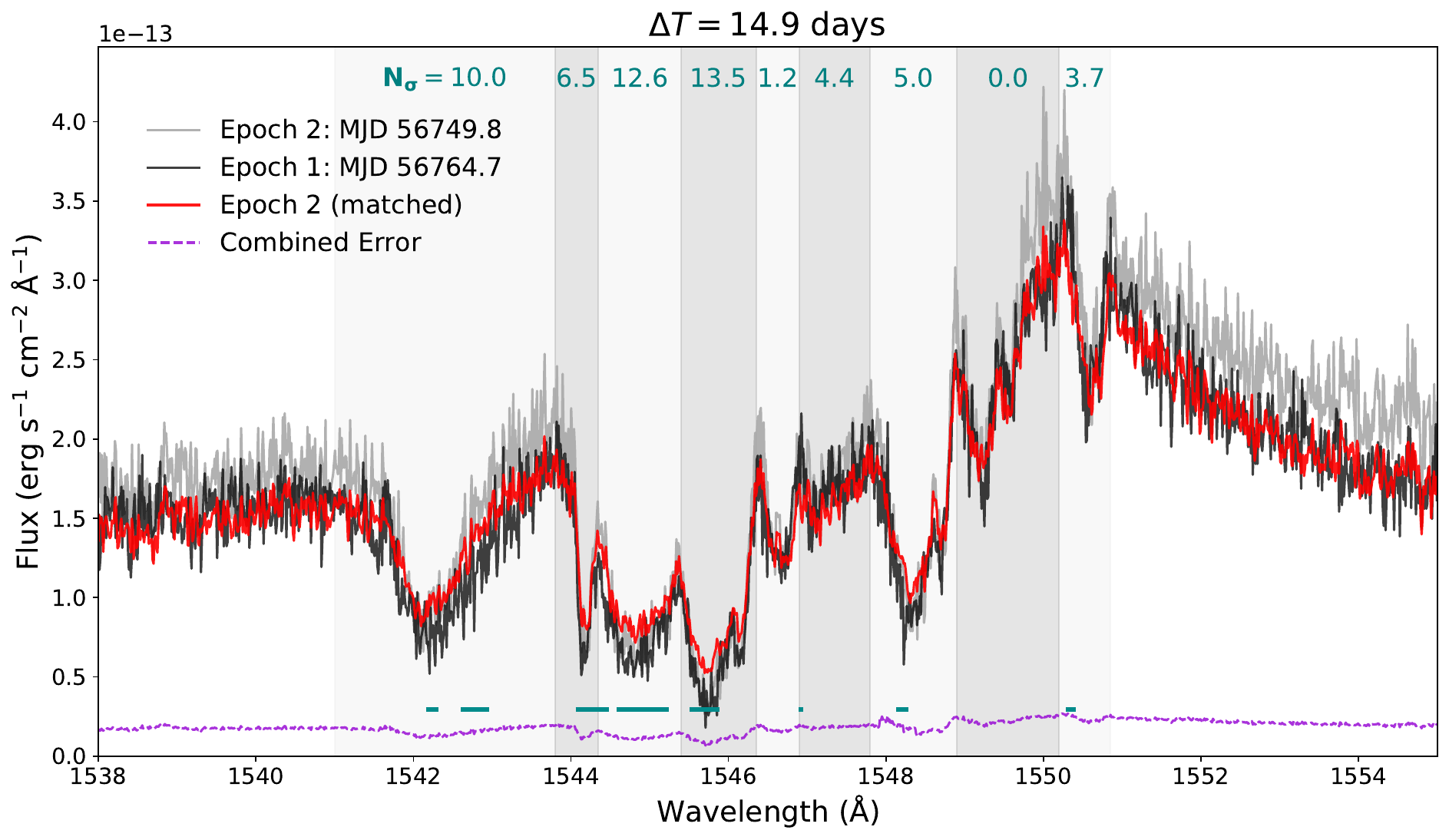}
    \caption{\textbf{Examples of absorption-line variability detection based on direct spectral comparison. }
    Each panel shows a pair of C\,\textsc{iv} absorption spectra taken at two epochs, with the time interval $\Delta T$ indicated in the title. The spectrum with higher signal-to-noise (Epoch 2, gray solid line) is used as the reference and is matched to the other epoch (Epoch 1, black solid line) by applying a double power-law scaling and adding two Gaussian components near the emission line to account for emission variability (one broad, one narrow). The matched spectrum (red solid line) is overplotted, and the magenta dashed line shows the combined 1$\sigma$ uncertainty. 
    Cyan horizontal bars mark the regions where significant variability is detected. Shaded vertical bands indicate the predefined absorption troughs used in variability analysis. Within each trough, the variability significance $N_\sigma$ is labeled at the top.
    }
    \label{fig:var_examples_short}
\end{figure}

\begin{figure}[htbp]
    \centering
    \includegraphics[width=0.95\textwidth]{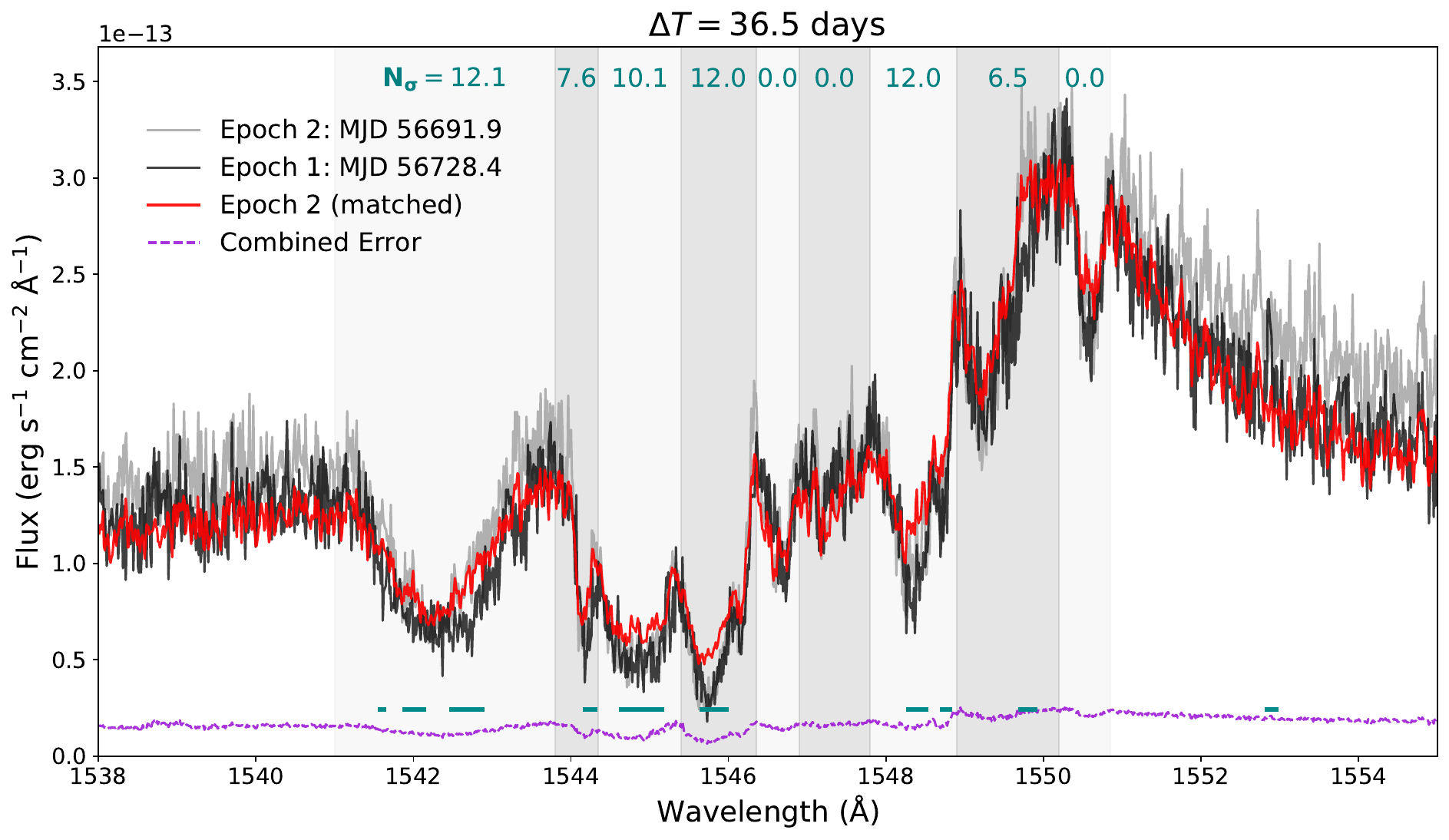}
    \includegraphics[width=0.95\textwidth]{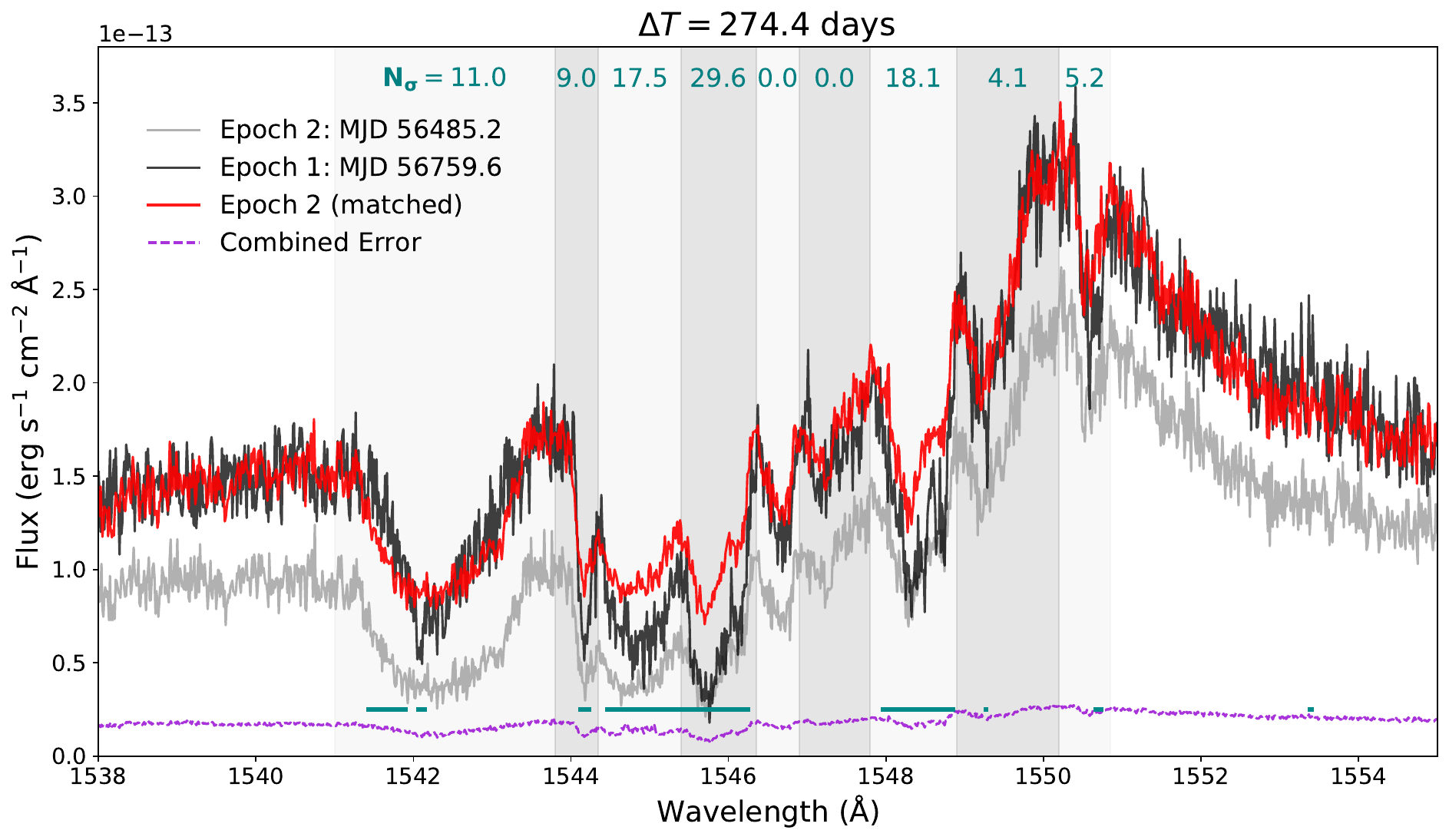}
    \caption{
    Same as Figure~\ref{fig:var_examples_short}, but for two long-term variability examples.
    }
    \label{fig:var_examples_long}
\end{figure}

\section{acknowledgments}

We acknowledge the research grants from the Ministry of Science and Technology of China (National Key Program for Science and Technology Research and Development, No. 2023YFA1608100), the research grants from the China Manned Space Project (CMS-CSST-2025-A08), the National Natural Science Foundation of China (Nos. 12273036, 12222304, 12192220 and 12192221).
This research is based on observations made with the NASA/ESA Hubble Space Telescope obtained from the Space Telescope Science Institute, which is operated by the Association of Universities for Research in Astronomy, Inc., under NASA contract NAS 5–26555. These observations are associated with programs GO-13330 and GO-13184. The HST data presented in this article were obtained from the Mikulski Archive for Space Telescopes (MAST) at the Space Telescope Science Institute. The specific observations analyzed can be accessed via \dataset[doi: 10.17909/mmy4-vz85]{https://doi.org/10.17909/mmy4-vz85}.

\appendix
\section{Ionization Parameter and Hydrogen Column Density}
\label{appendix:cloudy}
To model the ionization structure of the outflowing gas and predict ionic column densities, we employed version C23 of the spectral synthesis code CLOUDY (\cite{Cloudy2023}). We constructed a grid of CLOUDY simulations over a range of ionization parameter $U_H$ and hydrogen column density $N_H$. For each grid point, CLOUDY solves the photoionization and thermal equilibrium equations to compute the resulting ionic column densities. These model predictions were then compared to the observed column densities of C IV and Si IV, and the model with the best match was selected as the best-fit solution, thereby constraining the physical conditions of the absorbing gas.

Constraints on absorption properties come from the ionic column densities of C IV and Si IV, derived from the composite spectrum built by taking the geometric mean of the 76 observed spectra. 
We scaled the composite quasar template without absorption features (\cite{wang_2015}) using a double power-law function \eqref{eq:double-power-law}, and added Gaussian components at the emission-line positions to model emission-line variations, thereby obtaining the continuum plus emission-line flux of NGC 5548. By dividing the composite spectrum by the fitted result, we obtained the normalized flux.
For each absorption trough, the column densities were obtained by fitting the normalized flux using a partial covering absorption model. The normalized flux of either member of a resonance doublet is described by (\cite{1999ApJ...516...27A, 2002ApJS..141..267H}):
$$
I(\nu) = 1 - C(\nu) + C(\nu) e^{-\tau(\nu)}
$$
where $C(\nu)$ and $\tau(\nu)$ denote the covering factor and optical depth, respectively, both expressed as functions of velocity $\nu$.

The theoretical optical depth ratio for resonance doublets is determined by the ratio of oscillator strengths. For C IV, the oscillator strengths of the 1548.2 Å and 1550.8 Å transitions are $f_{\mathrm{blue}} = 0.19$ and $f_{\mathrm{red}} = 0.095$, respectively; for Si IV, the 1393.8 Å and 1402.8 Å lines have $f_{\mathrm{blue}} = 0.513$ and $f_{\mathrm{red}} = 0.255$. Thus, the theoretical optical depth ratio is approximately 2. In this study, we fixed the optical depth of the blue member to be twice that of the red.

In the fitting procedure, we constructed arrays of $\tau(\nu)$ and $C(\nu)$ with an interval of $dv = 15~\mathrm{km~s^{-1}}$. This interval was chosen to avoid overfitting caused by an excessive number of free parameters. The optical depth of the blue member was set to be twice that of the red, and both members shared the same $C(\nu)$. At the overlapping regions of the blue and red members, the relative flux was taken as the product of the two; in the non-overlapping regions, the relative flux was represented by the single member alone. The relative flux defined on the velocity grid was then linearly interpolated onto the wavelength grid and fitted to the observed normalized flux spectrum, yielding the optical depth. As shown in Figure \ref{fig:partial-covering-model-fit}.

With the optical depth profile $\tau(\nu)$ determined, the ionic column density was calculated using (\cite{1991ApJ...379..245S}):
$$
N_{\mathrm{ion}} = \frac{3.7679 \times 10^{14}\,\mathrm{cm}^{-2}}{\lambda f} \int \tau(\nu) \, d\nu,
$$
where $\lambda$ and $f$ denote the transition wavelength and oscillator strength, and the integration is performed over velocity space (in units of km s$^{-1}$).

The derived logarithmic column densities for C IV are $15.77 \pm 0.05$, $13.56 \pm 0.05$, $14.36 \pm 0.04$, $14.79 \pm 0.01$, $15.64 \pm 0.03$, and $14.70 \pm 0.12$ (in $\log\,\mathrm{cm}^{-2}$).
For Si IV, the corresponding values are $14.51 \pm 0.09$, $12.98 \pm 0.13$, $13.75 \pm 0.03$, $13.20 \pm 0.03$, $13.27 \pm 0.04$, and $11.44 \pm 5.86$.

The $(Log\ U_H, Log\ N_H)$ parameter grid in CLOUDY simulations was sampled with a step size of 0.1 dex. The spectral energy distribution (SED) model adopted in this study is D22 (\cite{D22}), which was constructed based on the 2014 observations from the AGN STORM project. Based on the average luminosity at 1500 Å from the composite spectrum, we estimated the bolometric luminosity to be $L_{\mathrm{bol}} = 2.01 \times 10^{44} \mathrm{erg} \mathrm{s}^{-1}$, which corresponds to a hydrogen-ionizing photon emission rate of $Q_H = 1.51 \times 10^{54} \mathrm{s}^{-1}$.

Given that AGN outflows typically exhibit super-solar metallicity, we adopted a metal abundance of $Z = 2Z_\odot$ in our simulations (\cite{2005ApJ...634..928F,2007ApJ...658..829A}). The best-fit models were determined by minimizing the following $\chi^2$ statistic between observed and modeled ionic column densities (\cite{Arav_2015}):
$$
\chi^2 = \sum_i \left( \frac{\log N_{i,\mathrm{mod}} - \log N_{i,\mathrm{obs}}}{\log N_{i,\mathrm{obs}} - \log (N_{i,\mathrm{obs}} \pm \sigma_i)} \right)^2,
$$
where $N_{i,\mathrm{mod}}$ and $N_{i,\mathrm{obs}}$ represent the modeled and observed column densities for ion species $i$, respectively, and $\sigma_i$ denotes the observational uncertainties. Positive uncertainties $+\sigma_i$ are used when $N_{i,\mathrm{mod}} > N_{i,\mathrm{obs}}$, and negative uncertainties $-\sigma_i$ otherwise, to account for asymmetry in the error bars. 

After obtaining the best-fit ionization parameter $U_H$ and hydrogen column density $N_H$, we used these values together with the other input parameters in Cloudy to simulate the number density ratio $\frac{n_{\mathrm{CV}}}{n_{\mathrm{CIV}}}$. The final simulation results are presented in Table \ref{table:D22} and Figure \ref{fig:cloudy_D22}.

\bibliography{refs}{}
\bibliographystyle{aasjournalv7}



\end{document}